\documentclass[amssymb,aps,prd,floatfix,twocolumn,nofootinbib]{revtex4}
\usepackage[intlimits]{amsmath}
\usepackage{amssymb}
\usepackage{graphicx}
\usepackage{tensor}
\usepackage[dvipsnames]{xcolor}
\usepackage{hyperref}
\usepackage{ulem} 

\begin{document}

\title{Monte Carlo methods for stationary solutions of general-relativistic Vlasov systems: Collisionless accretion onto black holes}

\author{Patryk Mach}
\affiliation{Instytut Fizyki Teoretycznej, Uniwersytet Jagiello\'{n}ski, \L ojasiewicza 11, 30-348 Krak\'ow, Poland}
\author{Adam Cie\'{s}lik}
\affiliation{Instytut Fizyki Teoretycznej, Uniwersytet Jagiello\'{n}ski, \L ojasiewicza 11, 30-348 Krak\'ow, Poland}
\author{Andrzej Odrzywo\l{}ek}
\affiliation{Instytut Fizyki Teoretycznej, Uniwersytet Jagiello\'{n}ski, \L ojasiewicza 11, 30-348 Krak\'ow, Poland}

\begin{abstract}
We develop a Monte Carlo simulation method for computing stationary solutions of the general-relativistic Vlasov equation describing a gas of non-colliding particles. As specific examples, we select planar or spherically symmetric accretion models on the Schwarzschild background spacetime. In all cases the gas extends to infinity, which poses an additional difficulty in the Monte Carlo approach. We discuss models with monoenergetic particles as well as solutions obeying the Maxwell-J\"{u}ttner distribution at infinity. For all models, exact expressions for the particle current density are known or can be computed analytically. We demonstrate perfect agreement between exact expressions for the particle current density and the results of our Monte Carlo simulations. 
\end{abstract}

\maketitle

\section{Introduction}

In the general-relativistic kinetic theory of gases observable quantities (the particle current density, the energy momentum tensor, the energy density, the pressure, etc.) are computed as suitable integrals over momentum space \cite{Synge, israel, Ehlers1, Ehlers2, Groot, Cercignani, AndreassonReview, zannias, acuna}. With a sufficient control of the phase-space structure (e.g., a good description of the regions in the phase space available for the motion of gas particles), one can, in many cases, compute such integrals directly, providing a solution to the problem at hand (see, e.g., \cite{Olivier1, Olivier2, cieslik, pmao1, pmao2, Gamboa, pmao3, cieslik_mach_odrzywolek_2022, pmao4}). For complex problems, a good description of the phase-space structure may not be available, and one has to resort to other methods. In this work we deal with a simple (but important) case in which collisions between individual particles of the gas are neglected. Since a collisionless gas consists of particles moving along geodesic trajectories, a possible way would be to construct a Monte Carlo simulation, in which one would select a sample of geodesic trajectories and then compute  averages over this sample, in order to obtain desired observables. Although this idea sounds simple, its actual implementation poses several difficulties related to the selection of geodesics and the averaging procedure.

Another option is to use the so-called particle-in-cell method, which has been employed successfully in general-relativistic simulations of kinetic systems with magnetic fields \cite{Bransgrove, Parfrey, Crinquand, Galishnikova}.

In this article we present a technical implementation of a Monte Carlo approach, designed to work with stationary solutions. While dealing with stationary solutions would seem to be natural and even simplifying, it requires a somewhat counterintuitive approach to the averaging process, as it will become clear from the remainder of this paper. In short, instead of counting intersections of geodesic trajectories with regions of spacelike hypersurfaces, we count intersections with suitable timelike hypersurfaces.

For clarity, we will focus on several easy accretion-type problems in the Schwarzschild spacetime, for which analytic solutions are available. Specifically, we will consider stationary Bondi-type solutions, in which the gas extends to infinity, where it is assumed to be homogeneous and at rest \cite{Shapiro, Olivier1, Olivier2}. We will work with planar models, assuming that the gas is confined to a single plane (similarly to the accretion model in the Kerr spacetime described in \cite{cieslik_mach_odrzywolek_2022}), and with spherically symmetric configurations. In the simplest case, we restrict ourselves to monoenergetic particles, but we also give examples in which the gas obeys the Maxwell-J\"{u}ttner distribution at infinity \cite{juttner1, juttner2}.

As a supplement to this paper, we provide sample Wolfram Mathematica \cite{Mathematica} codes performing our simulations. They will be publicly available at \cite{repo}.

The order of this paper is as follows. In Section \ref{sec:preliminaries} we introduce our notation and define the one-particle distribution function. Section \ref{sec:planar} is devoted to planar models, in which the motion of particles is confined to a common plane. We start with an elementary model of a planar uniform distribution of monoenergetic particles in the Minkowski spacetime. Two planar accretion models for the Schwarzschild spacetime are given in Sec.\ \ref{sec:planarschwarzschildexact}. In Section \ref{sec:montecarlointegration} we discuss Monte Carlo integration in the context of these two models. This part provides a link between the two approaches: an analytic one and an approach based on the actual Monte Carlo simulations; it also gives an opportunity to discuss the selection of geodesic parameters describing the Monte Carlo sample. Section \ref{sec:montecarlosimulations} introduces our Monte Carlo simulations. In Section \ref{sec:spherical} we deal with spherically symmetric stationary accretion models. A short summary is given in Sec. \ref{sec:summary}.

We will work in standard gravitational units with $c = G = 1$, where $c$ is the speed of light and $G$ denotes the gravitational constant. We assume the signature of the metric in the form $(-,+,+,+)$.

\section{One-particle distribution function}
\label{sec:preliminaries}

In the framework of the general-relativistic kinetic theory the gas is described in terms of a distribution function defined on a one-particle phase space---a subset of the tangent or cotangent bundle of the spacetime manifold (see, e.g., \cite{acuna} for a recent introduction). In this paper we adhere to the cotangent bundle formulation, but it is, in fact, a matter of convention.

Let $(M,g)$ be a spacetime manifold. The cotangent bundle of $M$ is defined as
\begin{equation}
T^\ast M = \{ (x,p) \colon x \in M, \, p \in T_x^\ast M \}.
\end{equation}
The one-particle distribution function is a real function defined on suitably chosen subsets $U \subseteq T^\ast M$, i.e., $\mathcal F \colon U \to [0,+\infty)$. The precise mathematical definition of $\mathcal F$ depends on a convention regarding the choice of $U$. Let us consider a gas composed of same rest-mass particles (such a gas is sometimes referred to as ``simple'' \cite{israel}). One possibility is to restrict $U$ to the future mass shell, defined as
\begin{eqnarray}
\Gamma^+_m & = & \{ (x,p) \in T^\ast M \colon g^{\mu \nu} p_\mu p_\nu = - m^2, \nonumber \\
&& p \text{ is future-directed} \},
\end{eqnarray}
where $m$ denotes the rest mass of a particle. This option has been chosen, e.g., in \cite{acuna}. Another option is to transfer the mass-shell restriction to the assumed form of $\mathcal F$, and to demand that $\mathcal F \sim \delta \left( \sqrt{-p_\mu p^\mu} - m \right)$. In this paper, we decided to choose the latter option, mainly to keep connection with previous works \cite{Olivier1,Olivier2}.

Let $S$ denote a 3-dimensional spacelike hypersurface in $M$, and let $s$ be a future-directed unit vector normal to $S$. We define $\mathcal N[S]$ as an averaged number of particle trajectories in $U$, whose projections on $M$ intersect $S$. In other words, $\mathcal N[S]$ denotes the number of particles in $S$. It can be shown that (cf.\ \cite{acuna}, Eqs.\ (94))
\begin{equation}
\label{defN}
\mathcal N[S] = - \int_S \left[ \int_{P_x^+}  \mathcal F(x,p) p_\mu s^\mu \mathrm{dvol}_x(p) \right] \eta_S,
\end{equation}
where
\begin{equation} P_x^+ = \{ p \in T_x^\ast M \colon g^{\mu \nu} p_\mu p_\nu < 0, \, p \text{ is future-directed} \}
\end{equation}
and $\eta_S$ denotes the volume element on $S$. The volume element on $P_x^+$ is given (in local adapted coordinates) by
\begin{equation} \mathrm{dvol}_x(p) = \sqrt{- \mathrm{det} \, g^{\mu \nu}(x)} dp_0 dp_1 dp_2 dp_3. \end{equation}

Equation (\ref{defN}) gives rise to the definition of the so-called particle current density
\begin{equation} \mathcal J_\mu(x) = \int_{P_x^+} \mathcal F (x,p) p_\mu \mathrm{dvol}_x(p). \end{equation}
With this definition one can write $\mathcal N[S]$ as
\begin{equation}
\label{NJ}
\mathcal N[S] = - \int_S \mathcal J_\mu s^\mu \eta_S. \end{equation}

The geodesic motion can be described by Hamilton's equations
\begin{subequations}
\label{hamiltonseqs}
\begin{eqnarray}
\frac{d x^\mu}{d \tau} & = & \frac{\partial H(x,p)}{\partial p_\mu}, \\
\frac{d p_\nu}{d \tau} & = & - \frac{\partial H(x,p)}{\partial x^\nu},
\end{eqnarray}
\end{subequations}
where $p^\mu = d x^\mu / d \tau$, $H(x,p) = \frac{1}{2} g^{\mu\nu}(x^\alpha) p_\mu p_\nu = - \frac{1}{2} m^2$. In the absence of collisions between particles, the distribution function $\mathcal F$ satisfies the so-called Vlasov equation, which can be expressed as a requirement that $\mathcal F$ should remain constant along a geodesic:
\begin{eqnarray}
\label{vlasoveq}
    \frac{d\mathcal F}{d\tau} & = & \frac{\partial \mathcal F}{\partial x^\mu} \frac{d x^\mu}{d \tau} + \frac{\partial \mathcal F}{\partial p_\nu} \frac{d p_\nu}{d \tau} = \frac{\partial \mathcal F}{\partial x^\mu} \frac{\partial H}{\partial p_\mu} - \frac{\partial \mathcal F}{\partial p_\nu} \frac{\partial H}{\partial x^\nu} \nonumber \\
    & = & \{ H, \mathcal F \} = 0.
\end{eqnarray}
Here $\{ \cdot, \cdot \}$ denotes the Poisson bracket. Note in particular that a probability function depending on $(x,p)$ via constants of motion $I_i(x,p)$, i.e., $\mathcal F = \mathcal F(I_1(x,p),\dots,I_s(x,p))$, where $\{H,I_i\} = 0$, $i = 1, \dots, s$, would always satisfy the Vlasov equation. In explicit terms, Eq.\ (\ref{vlasoveq}) can be written as
\begin{equation}
g^{\mu\nu} p_\nu \frac{\partial \mathcal F}{\partial x^\mu} - \frac{1}{2} p_\alpha p_\beta \frac{\partial g^{\alpha \beta}}{\partial x^\mu} \frac{\partial \mathcal F}{\partial p_\mu} = 0.
\end{equation}

Using Eq.\ (\ref{vlasoveq}), one can show that the particle current density satisfies the conservation law $\nabla_\mu \mathcal J^\mu = 0$, which again justifies formula (\ref{NJ}). The particle number density can be defined covariantly as $n = \sqrt{- \mathcal J_\mu \mathcal J^\mu}$. Alternatively, one can work with the components of $\mathcal J_\mu$.

\section{Particles confined to a plane}
\label{sec:planar}

We will assume a specific-to-general approach and start with illustrative cases of the gas confined to a plane. More specifically, we will begin with a discussion of a uniform gas of monoenergetic particles restricted to a two dimensional plane in the flat Minkowski spacetime. In the next step, we turn to a planar accretion problem in the Schwarzschild spacetime.

\subsection{Flat space case}
\label{sec:flatminkowski}

Consider a uniform gas of monoenergetic non-colliding particles confined to a two dimensional plane in the flat Minkowski spacetime. The gas particles move along straight lines. The only non vanishing component of the particle current density $\mathcal J_t$ can be computed in two equivalent ways. A straightforward way is to work in Cartesian coordinates $(t,x,y,z,)$. Defining
\begin{equation}
\mathcal F (x^\alpha,p_\beta) = \delta(z) F(t,x,y; p_\beta),
\end{equation}
we get
\begin{equation}
\mathcal J_\mu(x^\alpha) = \delta(z) J_\mu (t,x,y),
\end{equation}
where
\begin{equation}
J_\mu (t,x,y) = \int F (t,x,y;p_\beta) p_\mu \mathrm{dvol}_x(p).
\end{equation}

For a uniformly distributed gas of monoenergetic particles of the same mass $m_0$, we assume
\begin{eqnarray}
F(x,p) & = & \alpha m_0 \delta(p_z) \delta(p_t + E_0) \delta \left( \sqrt{p_t^2 - \mathbf p^2} - m_0 \right) \nonumber \\
& = & \alpha \delta(p_z) \delta(\varepsilon - \varepsilon_0) \delta \left( m - m_0 \right),
\end{eqnarray}
where $p_t = - E$, $E = m \varepsilon$, $E_0 = m \varepsilon_0$, and $\alpha$ is a proportionality constant. Hence
\begin{eqnarray}
J_t & = & \alpha  m_0 \int  \delta(p_z) \delta(p_t + E_0) \nonumber \\
&& \times \delta \left( \sqrt{p_t^2 - \mathbf p^2} - m_0 \right) p_t dp_t dp_x dp_y dp_z.
\end{eqnarray}
Integrating over $p_t$ and $p_z$ is straightforward and gives
\begin{equation}
J_t = - \alpha m_0 E_0 \int  \delta \left( \sqrt{E_0^2 - p_x^2 - p_y^2} - m_0 \right) dp_x dp_y.
\end{equation}
Introducing polar momentum coordinates $p_x = \zeta \cos \vartheta$, $p_y = \zeta \sin \vartheta$, we get $dp_x dp_y = \zeta  d \vartheta d \zeta$ and
\begin{equation}
J_t = - 2 \pi \alpha m_0 E_0 \int_0^\infty  \delta \left( \sqrt{p_t^2 - \zeta^2} - m_0 \right) \zeta d \zeta.
\end{equation}
On the other hand    
\begin{equation}
\delta \left( \sqrt{p_t^2 - \zeta^2} - m_0 \right) = \frac{m_0}{\sqrt{E_0^2 - m_0^2}} \delta \left( \zeta - \sqrt{E_0^2 - m_0^2} \right),
\end{equation}
and hence
\begin{equation}
\label{Jtflat}
J_t = - 2 \pi \alpha m_0^2 E_0 = - 2 \pi \alpha m_0^3 \varepsilon_0,
\end{equation}
where $E_0 = m_0 \varepsilon_0$.

The second calculation is based on spherical coordinates. While they may seem to make a very bad choice for the problem with an explicit translational symmetry, this calculation provides an illustration of some aspects related to the selection of samples of geodesics in the Monte Carlo simulations discussed in this paper. We define
\begin{subequations}
\begin{eqnarray}
\label{a20}
    m^2 & = &  p_t^2 - p_r^2 - \frac{1}{r^2} p_\theta^2 - \frac{1}{r^2 \sin^2 \theta} p_\varphi^2, \\
    E & = & - p_t, \\
    l^2 & = & p_\theta^2 + \frac{1}{\sin^2 \theta} p_\varphi^2, \\
    l_z & = & p_\varphi
\end{eqnarray}
\end{subequations}
and assume a convention in which $l \ge 0$. Quantities $m$, $E$, $l$, and $l_z$ are constants of motion. A straightforward calculation gives
\begin{equation}
\frac{\partial (m^2,E,l^2,l_z)}{\partial (p_t,p_r,p_\theta,p_\varphi)} = \pm 4 \sqrt{E^2 - m^2 - \frac{l^2}{r^2}} \sqrt{l^2 - \frac{l_z^2}{\sin^2 \theta}}.
\end{equation}
Treating $m$, $E$, $l$, and $l_z$ as new momentum coordinates, one obtains
\begin{eqnarray}
m l dm dE dl dl_z & = & \sqrt{E^2 - m^2 - \frac{l^2}{r^2}} \\
& & \times \sqrt{l^2 - \frac{l_z^2}{\sin^2 \theta}} dp_t dp_r dp_\theta dp_\varphi. \nonumber
\end{eqnarray}

Keeping the pair of coordinates $(l,l_z)$ is troublesome in making the restriction to the equatorial plane. A convenient solution is to change $(l,l_z) \mapsto (l,\sigma)$, where
\begin{equation}
\label{sigma}
\sin \sigma = \frac{l_z}{l}.
\end{equation}
Thus, $p_\theta = 0$ corresponds to $\sigma = \pm \pi/2$. We have $d l_z = \sqrt{l^2 - l_z^2} d \sigma$. Moreover, at the equatorial plane $\theta = \pi/2$, we get
\begin{equation}
m l dm dE dl d \sigma = \sqrt{E^2 - m^2 - \frac{l^2}{r^2}} dp_t dp_r dp_\theta dp_\varphi,
\end{equation}
and hence the volume element $\mathrm{dvol}_x(p)$ can be written as 
\begin{equation}
\mathrm{dvol}_x(p) = \frac{m l dm dE dl d \sigma }{r^2 \sqrt{E^2 - m^2 - \frac{l^2}{r^2}}}.
\end{equation}
Similarly,
\begin{equation}
\label{deltapz}
\delta(p_z) = \frac{r}{l} \left[ \delta \left(\sigma - \frac{\pi}{2} \right) + \delta \left( \sigma + \frac{\pi}{2} \right) \right].
\end{equation}
This gives
\begin{eqnarray}
F(x,p) & = & \frac{\alpha m_0 r}{l} \delta(m - m_0) \delta(E - E_0) \nonumber \\
& & \times \left[ \delta \left(\sigma - \frac{\pi}{2} \right) + \delta \left( \sigma + \frac{\pi}{2} \right) \right]
\end{eqnarray}
and
\begin{equation}
J_t = - \frac{4 \alpha m_0^2 E_0}{r} \int_0^{l_\mathrm{max}} \frac{dl}{\sqrt{E_0^2 - m_0^2 - \frac{l^2}{r^2}}},
\end{equation}
where $l_\mathrm{max} = r \sqrt{E_0^2 - m_0^2}$. The additional factor 2 appearing in the above formula comes from the fact that a given set of values $E$, $m$ and $l$ corresponds to 2 possible values of the radial momentum $p_r = \pm |p_r|$ (cf. Eq.\ (\ref{a20})). Let us introduce the following dimensionless quantities
\begin{equation}
l = M m_0 \lambda, \quad r = M \xi, \quad E_0 = m_0 \varepsilon_0.
\end{equation}
The expression for $J_t$ can be written as
\begin{equation}
\label{Jtmink1}
J_t = - \frac{4 \alpha m_0^3 \varepsilon_0}{\xi} \int_0^{\lambda_\mathrm{max}} \frac{d \lambda}{\sqrt{\varepsilon_0^2 - 1 - \frac{\lambda^2}{\xi^2}}},
\end{equation}
where $\lambda_\mathrm{max} = \xi \sqrt{\varepsilon_0^2 - 1}$. The integral in Eq.\ (\ref{Jtmink1}) reads
\begin{equation}
\label{Jtmink2}
\int_0^{\lambda_\mathrm{max}} \frac{d \lambda}{\sqrt{\varepsilon_0^2 - 1 - \frac{\lambda^2}{\xi^2}}} = \frac{\pi \xi}{2}.
\end{equation}
Note that it can also be expressed as
\begin{equation}
\label{Jtmink3}
\frac{\pi}{2 \sqrt{\varepsilon_0^2 - 1}} \int_0^{\lambda_\mathrm{max}} d \lambda = \frac{\pi \xi}{2}.
\end{equation}
Thus, we get finally
\begin{equation}
J_t = - 2 \pi \alpha m_0^3 \varepsilon_0,
\end{equation}
which coincides with Eq.\ (\ref{Jtflat}). We will return to Eqs.\ (\ref{Jtmink1}) and (\ref{Jtmink2}) in Sec.\ \ref{sec:montecarlosimulations}, discussing the selection of geodesic parameters. For a future use, note that
\begin{equation}
\label{ns}
n_s = 2 \pi \alpha m_0^3 \varepsilon_0
\end{equation}
can be interpreted as a particle surface density of the gas.

\subsection{Planar accretion onto a Schwarzschild black hole}
\label{sec:planarschwarzschildexact}

As one of our test models, we consider planar stationary accretion in the Schwarzschild spacetime. It is an equivalent of the Bondi (or Michel) type stationary accretion of the collisionless gas occurring in the Schwarzschild spacetime (see, \cite{Shapiro, Olivier1, Olivier2}), restricted to a plane. Within the plane, the gas extends to infinity, and it is assumed to be asymptotically ``at rest''. A model of this type for the Kerr spacetime has been analyzed in \cite{cieslik_mach_odrzywolek_2022}. In contrast to the Kerr case, the planar accretion model in the Schwarzschild spacetime is relatively simple and can serve as a pedagogical example in our discussion.

We work in standard Schwarzschild coordinates $(t,r,\theta,\varphi)$. The Schwarzschild metric has the well-known form
\begin{eqnarray}
    g & = & - \left( 1 - \frac{2 M}{r} \right) dt^2 + \left( 1 - \frac{2 M}{r}\right)^{-1} dr^2 \nonumber \\
    && + r^2 d \theta^2 + r^2 \sin^2 \theta d \varphi^2.
\end{eqnarray}
All our calculations can be also repeated using suitably chosen horizon-penetrating Eddington-Finkelstein type coordinates, yielding essentially the same results. We use the simplest Schwarzschild form of the metric to avoid unnecessary complications.

The geodesic motion is integrable. There exist 4 independent constants of motion, which for simplicity, we denote as
\begin{subequations}
\begin{eqnarray}
    m^2 & = & - g^{\mu \nu} p_\mu p_\nu, \\
    E & = & - p_t, \\
    l^2 & = & p_\theta^2 + \frac{1}{\sin^2 \theta} p_\varphi^2, \\
    l_z & = & p_\varphi,
\end{eqnarray}
\end{subequations}
using the symbols, which have already been introduced in the flat Minkowski case. As before, we assume $l \ge 0$ and define dimensionless quantities $\xi$, $\varepsilon$, $\lambda$, and $\lambda_z$:
\begin{equation}
r = M \xi, \quad E = m \varepsilon, \quad l_z = M m \lambda_z, \quad l = M m \lambda,
\end{equation}
where this time $M$ refers explicitly to the Schwarzschild mass. We will also use the momentum-space coordinate $\sigma$ \eqref{sigma}, defined using dimensionless quantities by $\sin \sigma = \lambda_z/\lambda$.

The phase-space region available for the motion can be controlled with almost the same expressions as those used in the spherically symmetric model \cite{Olivier1}.  In what follows, we will summarize most important formulas; details of the derivation can be found in \cite{Olivier1}. The contravariant radial component of the momentum vector reads
\begin{equation}
p^r = \pm m \sqrt{\varepsilon^2 - U_\lambda(\xi)},
\end{equation}
where
\begin{equation}
U_\lambda(\xi) = \left( 1 - \frac{2}{\xi} \right) \left( 1 + \frac{\lambda^2}{\xi^2} \right)
\end{equation}
denotes the dimensionless effective radial potential.

The region available for the radial motion is characterized by the condition $\varepsilon^2 - U_\lambda(\xi) \ge 0$. Since $U_\lambda(\xi) \to 1$, as $\xi \to \infty$, unbound orbits are characterized by the energy $\varepsilon \ge 1$. The orbits with the angular momentum $\lambda < \lambda_c(\varepsilon)$, where
\begin{equation}
    \lambda_c(\varepsilon)^2 = \frac{12}{1-\frac{4}{\left(\frac{3 \varepsilon }{\sqrt{9 \varepsilon ^2-8}}+1\right)^2}},
    \end{equation}
plunge into the black hole---we refer to these orbits as absorbed ones. Unbound orbits with $\lambda > \lambda_c(\varepsilon)$ are scattered off the centrifugal barrier. The maximum value of the angular momentum for a scattered unbound orbit reaching radius $\xi$ reads
    \begin{equation}
    \lambda_\mathrm{max} (\varepsilon, \xi) = \xi \sqrt{\frac{\varepsilon^2}{1 - \frac{2}{\xi}} - 1},
\end{equation}
while the minimum allowed energy can be written as
\begin{equation}
\label{limit2}
    \varepsilon_\mathrm{min}(\xi) = \begin{cases} \infty, & \xi \leq 3 ,\\
    \sqrt{\left(1 - \frac{2}{\xi} \right) \left(1 + \frac{1}{\xi - 3} \right)}, & 3 < \xi < 4, \\
    1, & \xi \ge 4.  \end{cases}
\end{equation}

\subsubsection{Monoenergetic particles}

For a planar accretion model with monoenergetic particles we assume the one-particle distribution function of the form
\begin{eqnarray}
F(x, p) & = & \alpha m_0 \delta\left(\sqrt{- p_\mu p^\mu} - m_0\right) \delta(p_t + E_0) \delta(p_z) \nonumber \\
& = & \alpha \delta\left(m - m_0\right)  \delta(\varepsilon - \varepsilon_0) \delta(p_z) \nonumber \\
& = & \frac{\alpha \xi}{m_0 \lambda} \delta\left(m - m_0 \right)  \delta(\varepsilon - \varepsilon_0) \nonumber \\
&& \times \left[ \delta(\sigma - \pi/2) + \delta(\sigma + \pi/2) \right].
\end{eqnarray}
One may readily verify that the above choice satisfies the Vlasov equation (\ref{vlasoveq}).

The volume element in the momentum space reads, in terms of coordinates $m$, $\varepsilon$, $\lambda$, and $\sigma$,
\begin{equation}
\mathrm{dvol}_x(p) = \frac{m^3 \lambda}{\xi^2 \sqrt{\varepsilon^2 - U_\lambda(\xi)}} dm d\varepsilon d\lambda d\sigma.
\end{equation}

The components of the particle surface current density $J_\mu$ can be now expressed as a sum of two parts: a part corresponding to absorbed orbits and a part corresponding to scattered ones, i.e., $J_\mu = J_\mu^\mathrm{(abs)} + J_\mu^\mathrm{(scat)}$. According to the characterization given in the previous subsection, one has
\begin{widetext}
\begin{subequations}
\label{Jtmonoplanar}
\begin{eqnarray}
    J_t^\mathrm{(abs)}(\xi) & = & - \frac{2 \alpha m_0^3}{\xi} \int_1^\infty d \varepsilon \delta(\varepsilon - \varepsilon_0) \varepsilon  \int_0^{\lambda_c(\varepsilon)} \frac{d \lambda}{\sqrt{\varepsilon^2 - U_\lambda (\xi)}} = - \frac{2 \alpha m_0^3}{\xi} \varepsilon_0 \Theta(\varepsilon_0  - 1) \int_0^{\lambda_c(\varepsilon_0)} \frac{d \lambda}{\sqrt{\varepsilon_0^2 - U_\lambda (\xi)}}, \\
    J_t^\mathrm{(scat)}(\xi) & = & - \frac{4 \alpha m_0^3}{\xi} \int_{\varepsilon_\mathrm{min}(\xi)}^\infty d \varepsilon \delta(\varepsilon - \varepsilon_0) \varepsilon \int_{\lambda_c(\varepsilon)}^{\lambda_\mathrm{max}(\varepsilon,\xi)} \frac{d \lambda}{\sqrt{\varepsilon^2 - U_\lambda (\xi)}} \nonumber \\
    & = & - \frac{4 \alpha m_0^3}{\xi} \varepsilon_0 \Theta(\varepsilon_0  - \varepsilon_\mathrm{min}(\xi)) \int_{\lambda_c(\varepsilon_0)}^{\lambda_\mathrm{max}(\varepsilon_0,\xi)} \frac{d \lambda}{\sqrt{\varepsilon_0^2 - U_\lambda (\xi)}},
\end{eqnarray}
\end{subequations}
\end{widetext}
where $\Theta$ denotes the Heaviside step function. The additional factor 2 in the expression for $J_t^\mathrm{(scat)}$ is due to equal contributions of ingoing and outgoing trajectories. The radial component reads
\begin{equation}
\label{Jrmonoplanar}
J^r(\xi) = J^r_\mathrm{(abs)}(\xi) = - \frac{2 \alpha m_0^3}{\xi} \Theta(\varepsilon_0 - 1) \lambda_c(\varepsilon_0).
\end{equation}
The radial component $J^r$ is directly related with the rest-mass accretion rate, which we define as (see \cite{cieslik_mach_odrzywolek_2022})
\begin{equation}
\dot M = - 2 \pi M m_0 \xi J^r.
\end{equation}
For the monoenergetic planar model the rest-mass accretion rate reads
\begin{equation}
\label{dotm}
\dot M = 4 \pi \alpha M m_0^4 \Theta(\varepsilon_0 - 1) \lambda_c(\varepsilon_0).
\end{equation}

As a general remark, applying to all models in this paper, let us note that in physical applications the proportionality constant $\alpha$ appearing in the expression for the distribution function is rather hard to control. For planar models, it can be expressed in terms of the asymptotic particle number surface density $n_s$, which for the gas of monoenergetic particles is given by Eq.\ (\ref{ns}). This allows us to write Eq.\ (\ref{dotm}) as
\begin{equation}
\dot M = 2 M m_0 n_s \Theta(\varepsilon_0 - 1) \frac{\lambda_c(\varepsilon_0)}{\varepsilon_0}.
\end{equation}
For spherical models one can express the proportionality constant $\alpha$ in terms of the asymptotic particle number density \cite{Olivier1, Olivier2}. Since in this work we focus mainly on the Monte Carlo simulations, we will keep the proportionality constant $\alpha$ for clarity of the resulting formulas.

\subsubsection{Maxwell-J\"{u}ttner distribution}

Assuming, instead of monoenergetic particles, that the gas obeys at infinity a Maxwell-J\"{u}ttner distribution restricted to a plane \cite{juttner1, juttner2}, we set
\begin{eqnarray}
F(x, p) & = & \alpha \delta\left(\sqrt{- p_\mu p^\mu} - m_0\right) \exp \left( \frac{\beta}{m_0} p_t \right) \delta(p_z) \nonumber \\
& = & \frac{\alpha \xi}{m_0 \lambda} \delta\left(m - m_0 \right)  \exp \left( -\beta \varepsilon \right) \nonumber \\
&& \times \left[ \delta(\sigma - \pi/2) + \delta(\sigma + \pi/2) \right],
\end{eqnarray}
where $\beta = m_0/(k_\mathrm{B} T)$, $k_\mathrm{B}$ denotes the Boltzmann constant, and $T$ is the asymptotic temperature of the gas.

A calculation similar to that for the planar accretion of monoenergetic particles yields
\begin{widetext}
\begin{subequations}
\label{Jexpplanar}
\begin{eqnarray}
    J_t^\mathrm{(abs)} (\xi) & = & - \frac{2 \alpha m_0^3}{\xi} \int_1^\infty \varepsilon  \exp(-\beta \varepsilon) \int_0^{\lambda_c(\varepsilon)} \frac{d \lambda}{\sqrt{\varepsilon^2 - U_\lambda(\xi)}}, \label{Jtabsexpplanar} \\
    J_t^\mathrm{(scat)} (\xi) & = & - \frac{4 \alpha m_0^3}{\xi} \int_{\varepsilon_\mathrm{min}(\xi)}^\infty \varepsilon  \exp(-\beta \varepsilon) \int_{\lambda_c(\varepsilon)}^{\lambda_\mathrm{max}(\varepsilon,\xi)} \frac{d \lambda}{\sqrt{\varepsilon^2 - U_\lambda(\xi)}}, \label{Jtscatexpplanar} \\
    J^r (\xi) & = & - \frac{2 \alpha m_0^3}{\xi} \int_1^\infty d \varepsilon \exp (- \beta \varepsilon) \lambda_c(\varepsilon). \label{Jrexpplanar}
\end{eqnarray}
\end{subequations}
\end{widetext}

With a slight abuse of terminology, we will refer to the model described in this subsection as the planar Maxwell-J\"{u}ttner model, keeping in mind that the Maxwell-J\"{u}ttner distribution is only assumed asymptotically.

\subsection{Monte Carlo integration}
\label{sec:montecarlointegration}

Our first approach is to apply a Monte Carlo integration to Eqs.\ (\ref{Jtmonoplanar}) or (\ref{Jexpplanar}) (for a general introduction to Monte Carlo methods see, for instance, \cite{kalos}). This procedure can be viewed as an intermediate step between analytic solutions given in  previous subsections and our final aim to construct actual Monte Carlo simulations. We will also use it as an illustration of the problems related with the selection of the sample of geodesics. Consider an integral
\begin{equation}
I = \int_\Omega f(x) dx.
\end{equation}
Let $X_i \in \Omega$, $i = 1, \dots N$ denote $N$ samples chosen from the distribution $p(x)$. The Monte Carlo estimator of $I$ reads
\begin{equation}
\langle I \rangle = \frac{1}{N} \sum_{i=1}^N \frac{f(X_i)}{p(X_i)}.
\end{equation}
For a uniform distribution, one has $p(x) = 1/V$, where $V = \mathrm{vol}(\Omega)$ is the volume of $\Omega$, and thus
\begin{equation}
\langle I \rangle = \frac{V}{N} \sum_{i=1}^N f(X_i). \end{equation}

Consider now an integral of the form
\begin{equation}
I = \int_\Omega f(x) g(x) dx,
\end{equation}
where $g(x)$ is a weight. For $p(x) = c g(x)$, where $c$ is a proportionality constant, one gets
\begin{equation}
\langle I \rangle = \frac{1}{N} \sum_{i=1}^N \frac{f(X_i) g(X_i)}{p(X_i)} = \frac{1}{c N} \sum_{i=1}^N f(X_i).
\end{equation}
Since the probability density function $p(x)$ integrates to unity, we get
\begin{equation}
c = \left[ \int_\Omega g(x) dx \right]^{-1}.
\end{equation}
Thus,
\begin{equation}
\langle I \rangle = \frac{1}{N} \left[ \int_\Omega g(x) dx \right] \sum_{i=1}^N f(X_i).
\end{equation}

\begin{figure}
    \centering
    \includegraphics[width=\columnwidth]{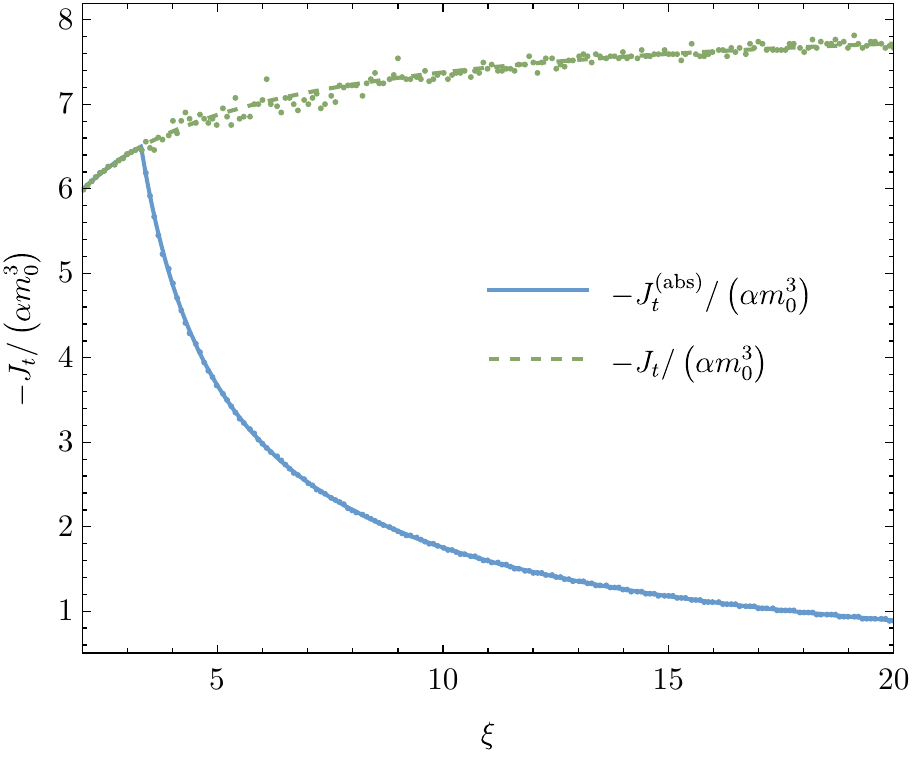}
    \caption{Components of the particle surface density current $J_t$ for the monoenergetic planar model with $\varepsilon_0 = 1.3$, $\xi_0 = 20$. Exact solutions (Eqs.\ (\ref{Jtmonoplanar}))are plotted with solid and dashed lines. Dots (blue and green) represent sample results obtained by the Monte Carlo integration (Eqs.\ (\ref{Jtmonointegration})). There are $10^5$ angular momentum samples: $N_\mathrm{abs} = 31814$, $N_\mathrm{scat} = 68186$.}
    \label{fig:my_label0a}
\end{figure}

These two options yield two different possibilities of attacking our problem. For monoenergetic accretion, we would rather use a uniform-distribution integration. This can be done as follows. Let $\xi_0$ be the outer radius of the region of interest, and let $\lambda_i \in [0,\lambda_\mathrm{max}(\varepsilon_0,\xi_0)]$ denote uniformly selected random angular momenta. Note that the configuration is assumed to extend to infinity. Selecting a finite $\xi_0$ allows us to focus attention on the region $2 < \xi \le \xi_0$ and to limit the range of the angular momentum.\footnote{Accretion from a region of a finite radius is an interesting problem on its own \cite{Gamboa}.} Angular momenta $\lambda_i$ can be divided into two classes with the corresponding sets of indices: $I_\mathrm{abs} = \{ i \colon 0 \le \lambda_i < \lambda_c(\varepsilon_0) \}$, $I_\mathrm{scat} = \{ i \colon \lambda_c < \lambda_i \le \lambda_\mathrm{max}(\varepsilon_0,\xi_0) \}$, with $N_\mathrm{abs} = \# I_\mathrm{abs}$, $N_\mathrm{scat} = \# I_\mathrm{scat}$, where $\# I$ denotes the number of elements in the set $I$. Monte Carlo estimators of $J_t^\mathrm{(abs)}$ and $J_t^\mathrm{(scat)}$ in the monoenergetic planar model can be written as
\begin{subequations}
\label{Jtmonointegration}
\begin{eqnarray}
    \langle J_t^\mathrm{(abs)} \rangle & = & - \frac{2 \alpha m_0^3 \lambda_c(\varepsilon_0)}{N_\mathrm{abs} \xi} \\
    && \times \sum_{i \in I_\mathrm{abs}}\frac{\varepsilon_0}{\sqrt{\varepsilon_0^2 - \left(1 - \frac{2}{\xi}\right)\left(1 + \frac{\lambda_i^2}{\xi^2}\right)}}, \nonumber \\
    \langle J_t^\mathrm{(scat)} \rangle & = & - \frac{4 \alpha m_0^3 [\lambda_\mathrm{max}(\varepsilon_0,\xi_0)  - \lambda_c(\varepsilon_0)]}{N_\mathrm{scat} \xi} \\
    && \times \sum_{i \in I_\mathrm{scat}} \frac{\varepsilon_0 \Theta(\xi - \xi_\mathrm{min}(\lambda_i,\varepsilon_0))}{\sqrt{\varepsilon_0^2 - \left(1 - \frac{2}{\xi}\right)\left(1 + \frac{\lambda_i^2}{\xi^2}\right)}}, \nonumber
\end{eqnarray}
\end{subequations}
where $\xi_\mathrm{min}(\lambda,\varepsilon)$ denotes the pericenter radius---the largest zero of the polynomial $\xi^3 \left[ \varepsilon^2 - U_\lambda(\xi) \right]$. Sample result obtained in this way are shown in Fig.\ \ref{fig:my_label0a}, assuming $\varepsilon_0 = 1.3$ and $\xi_0 = 20$. Monte Carlo data are plotted with dots. Solid and dashed lines represent exact solutions. We plot the graphs of $-J_t/\left( \alpha m_0^3 \right)$ and $-J_t^\mathrm{(abs)}/\left( \alpha m_0^3 \right)$. We omit the calculation of the Monte Carlo estimator for $J^r$, since it is almost trivial.

\begin{figure}
    \centering
    \includegraphics[width=\columnwidth]{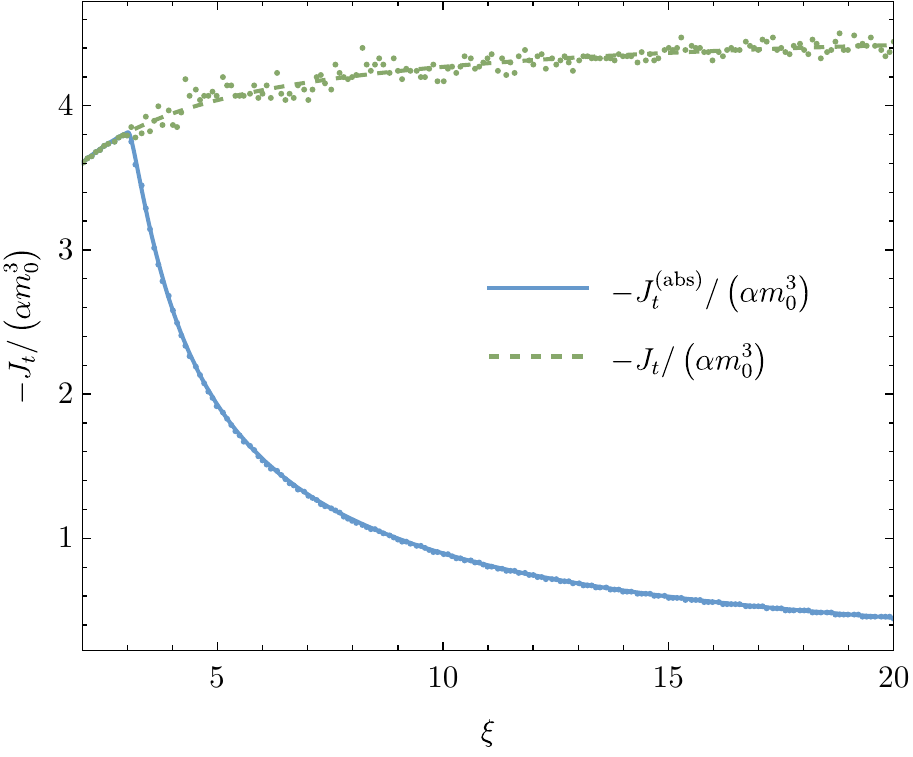}
    \caption{Same as in Fig.\ \ref{fig:my_label0a} but for the Maxwell-J\"{u}ttner planar accretion model with $\beta = 1$, $\varepsilon_\mathrm{cutoff} = 10$, and $\xi_0 = 20$. Exact solutions (Eqs.\ (\ref{Jtabsexpplanar}) and (\ref{Jtscatexpplanar})) are plotted with solid and dashed lines. Blue and green dots show results of the Monte Carlo integration (Eqs.\ \ref{Jtexpintegration}). The total number of randomly selected triples $(\varepsilon_i,\lambda_i,y_i)$ is $2 \times 10^6$, which gives $N_\mathrm{abs} = 10462$, $N_\mathrm{abs} = 27268$.}
    \label{fig:my_label0b}
\end{figure}

The planar Maxwell-J\"{u}ttner accretion model can serve as an illustration for the second scheme. Selecting geodesic parameters form the Maxwell-J\"{u}ttner distribution can be done using von Neumann's rejection method \cite{vonneumann}. As before, let us focus on the region $2 < \xi \le \xi_0$. We start by introducing a cutoff $\varepsilon_\mathrm{cutoff}$ for the allowed energy $\varepsilon$. Subsequently, we select, assuming uniform distributions, $\varepsilon_i \in [1, \varepsilon_\mathrm{cutoff}]$, $\lambda_i \in [0, \lambda_\mathrm{max}(\varepsilon_\mathrm{cutoff},\xi_0)]$, and an auxiliary variable $y_i \in [0,1]$. The values of $\varepsilon_i$ and $\lambda_i$ are added to the set of selected parameters, if $y_i < \exp (- \beta \varepsilon_i)/\exp(-\beta)$ and $\lambda_i < \lambda_\mathrm{max}(\varepsilon_i,\xi_0)$, and rejected otherwise. This procedure is iterated until a sufficient number of pairs $(\varepsilon_i, \lambda_i)$ is selected. As before, parameters $(\varepsilon_i, \lambda_i)$ are divided into  those corresponding to absorbed trajectories ($\lambda_i < \lambda_c(\varepsilon_i)$) and scattered ones ($\lambda_i > \lambda_c(\varepsilon_i)$). We denote, also as before, $N_\mathrm{abs} = \# I_\mathrm{abs}$, $N_\mathrm{scat} = \# I_\mathrm{scat}$.

Monte Carlo estimators of $J_t^\mathrm{(abs)}$ and $J_t^\mathrm{(scat)}$ are computed as
\begin{subequations}
\label{Jtexpintegration}
\begin{eqnarray}
    \langle J_t^\mathrm{(abs)} \rangle & = & - \frac{2 \alpha m_0^3 V_\mathrm{abs}}{N_\mathrm{abs} \xi}  \\
    && \times \sum_{i \in I_\mathrm{abs}}\frac{\varepsilon_i}{\sqrt{\varepsilon_i^2 - \left(1 - \frac{2}{\xi}\right)\left(1 + \frac{\lambda_i^2}{\xi^2}\right)}}, \nonumber \\
    \langle J_t^\mathrm{(scat)} \rangle & = & - \frac{4 \alpha m_0^3 V_\mathrm{scat}}{N_\mathrm{scat} \xi} \\
    && \times \sum_{i \in I_\mathrm{scat}} \frac{\varepsilon_i \Theta(\xi - \xi_\mathrm{min}(\lambda_i,\varepsilon_i))}{\sqrt{\varepsilon_i^2 - \left(1 - \frac{2}{\xi}\right)\left(1 + \frac{\lambda_i^2}{\xi^2}\right)}}, \nonumber
\end{eqnarray}
\end{subequations}
where
\begin{subequations}
\label{volumes}
\begin{eqnarray}
V_\mathrm{abs} & = & \int_1^{\varepsilon_\mathrm{cutoff}} \exp(- \beta \varepsilon) \lambda_c(\varepsilon) d \varepsilon, \\
V_\mathrm{scat} & = & \int_1^{\varepsilon_\mathrm{cutoff}} \exp(- \beta \varepsilon) [\lambda_\mathrm{max}(\varepsilon,\xi_0) - \lambda_c(\varepsilon)] d \varepsilon.
\end{eqnarray}
\end{subequations}
An example of the components $J_t^\mathrm{(abs)}$ and $J_t$ obtained in this way for $\beta = 1$, $\varepsilon_\mathrm{cutoff} = 10$, and $\xi_0 = 20$ is shown in Fig.\ \ref{fig:my_label0b}. In Figure \ref{fig:my_label0b} and in what follows, to assure a fair comparison, both analytic solutions and Monte Carlo estimators are computed assuming the same cutoff value $\varepsilon_\mathrm{cutoff}$ for the maximal energy.

In both cases we have introduced small adjustments in the Monte Carlo integration procedure, which make it similar to the actual Monte Carlo simulations discussed in the remainder of this paper. In particular, instead of limiting the selection of geodesic parameters used to compute $\langle J_t^\mathrm{(scat)} \rangle$ with $\lambda_\mathrm{max}(\varepsilon,\xi)$, we use an equivalent restriction on $\xi_\mathrm{min}$ --- the pericenter radius.

\subsection{Monte Carlo simulations of stationary planar problems}
\label{sec:montecarlosimulations}

\begin{figure}
    \centering
    \includegraphics[width=\columnwidth]{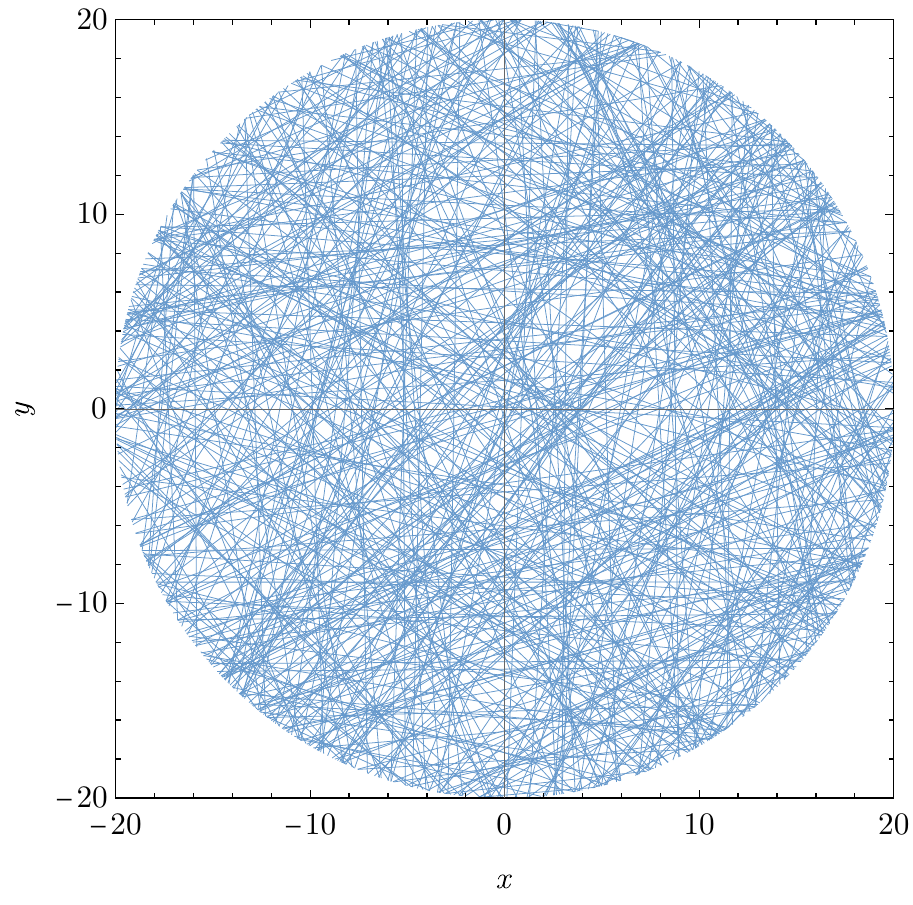}
    \caption{Randomly distributed straight lines on a plane. The lines in this picture are selected according to Eq.\ (\ref{poincare1}) or (\ref{poincare2}). There are 500 lines in this picture.}
    \label{fig:my_label1}
\end{figure}

\begin{figure}
    \centering
    \includegraphics[width=\columnwidth]{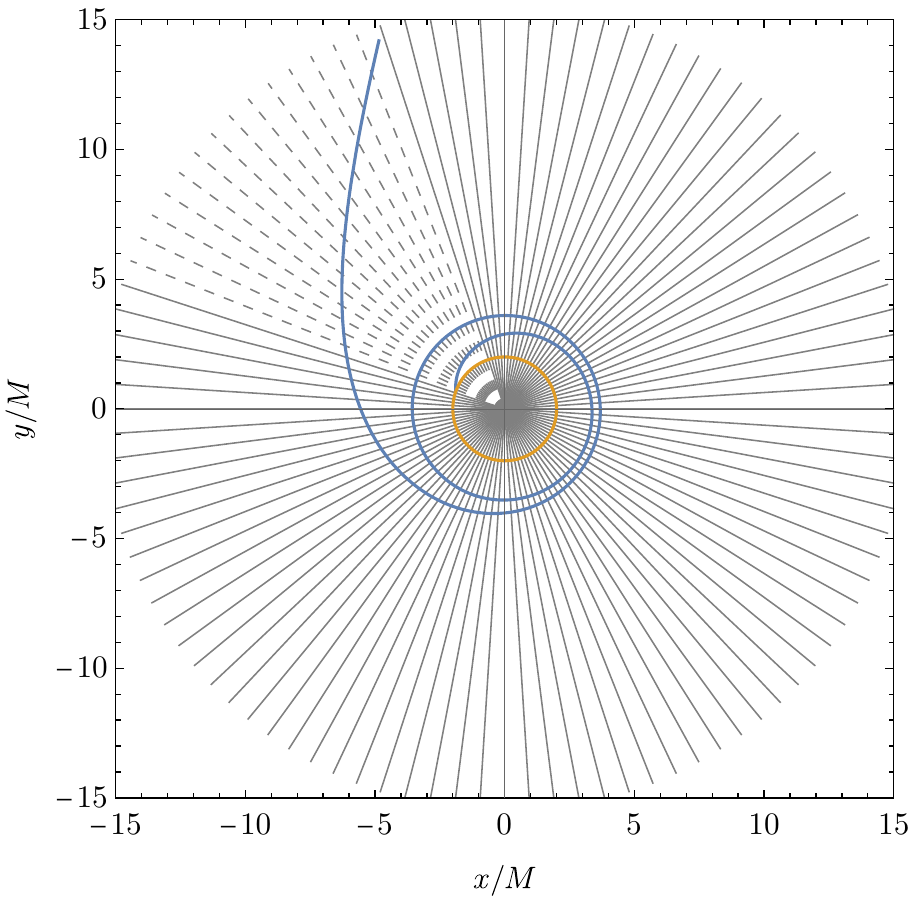}
    \caption{An illustration of the function controlling the number of intersections of a geodesic and a radial segment of constant $\varphi$. A spiraling geodesic is plotted with the blue line. It corresponds to $\varepsilon = 1.1$ and $\lambda = 4.7244$. Radial segments of constant $\varphi$ which are intersected twice are depicted with solid lines. A dashed line is used for rays intersected 3 times. The orange circle corresponds to the black hole horizon.}
    \label{fig:my_label2}
\end{figure}

\begin{figure}
    \centering
    \includegraphics[width=\columnwidth]{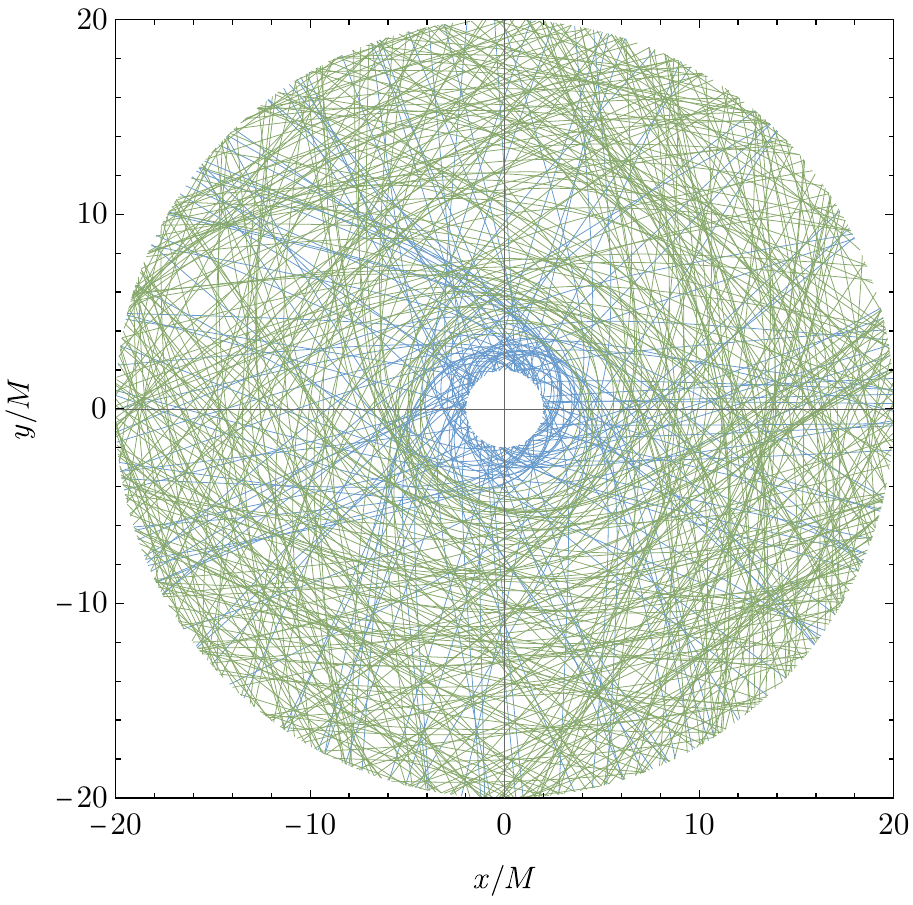}
    \caption{Randomly distributed monoenergetic ($\varepsilon_0 = 1.3$) geodesics on a plane. Angular momenta are uniformly distributed within $\lambda \in [0,\lambda_\mathrm{max}(\varepsilon_0,\xi_0)]$, where $\xi_0 = 20$. Absorbed orbits are depicted in blue. Scattered orbits are plotted in green. There are 500 orbits in this plot.}
    \label{fig:my_label3}
\end{figure}

\begin{figure}
    \centering
    \includegraphics[width=\columnwidth]{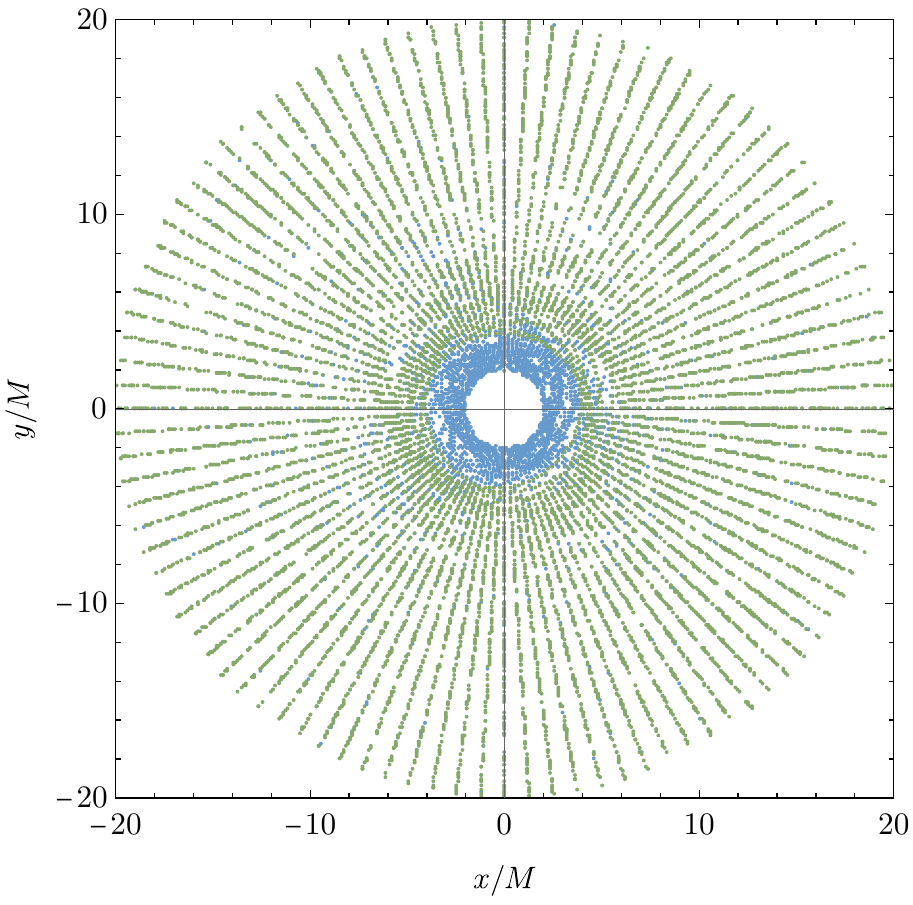}
    \caption{Intersection points of orbits plotted in Fig.\ \ref{fig:my_label3} and radii of constant $\varphi$.}
    \label{fig:my_label4}
\end{figure}

\begin{figure}
    \centering
    \includegraphics[width=\columnwidth]{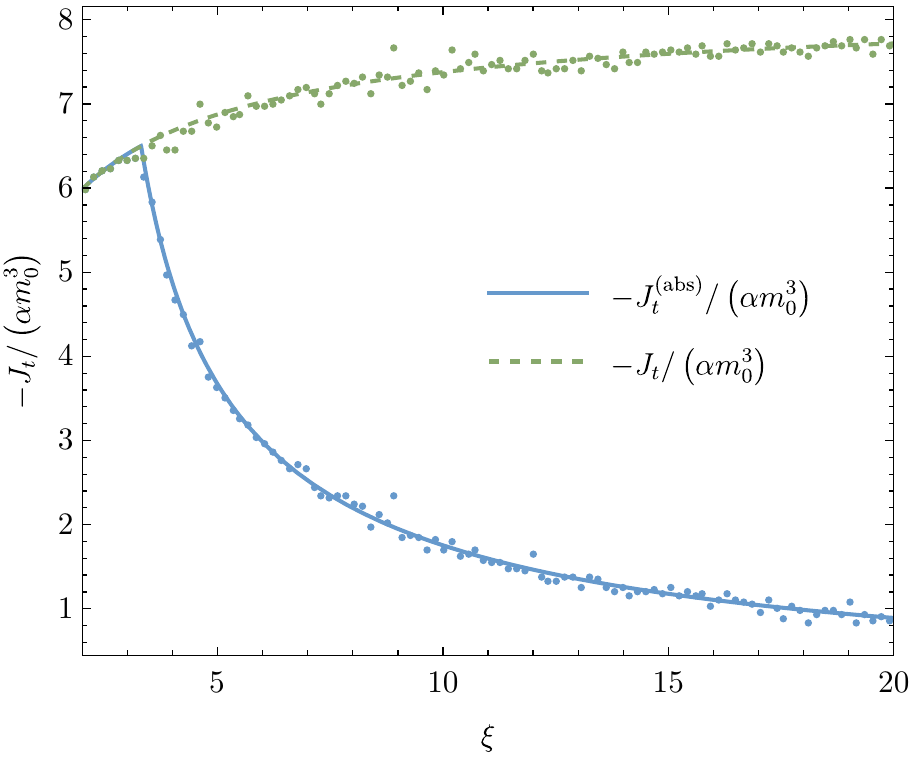}
    \caption{Time components of the particle surface density current $J_t$ for the planar accretion model with monoenergetic particles. In this case $\varepsilon_0 = 1.3$ and $\xi_0 = 20$. Exact solutions (Eqs.\ \ref{Jtmonoplanar}) are plotted with solid and dashed lines. Blue and green points depict results of a Monte Carlo simulation (Eqs.\ (\ref{JtMC2Dmono})). There are $5 \times 10^4$ orbits: 15953 absorbed trajectories and 34047 scattered ones.}
    \label{fig:my_label5}
\end{figure}

\begin{figure}
    \centering
    \includegraphics[width=\columnwidth]{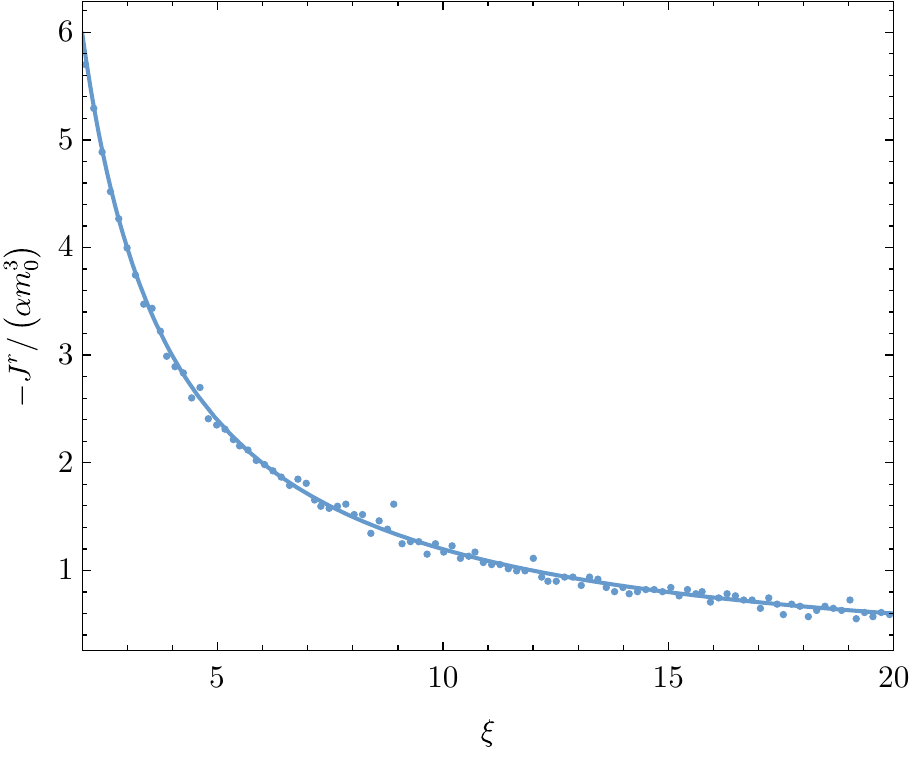}
    \caption{The radial component of the particle surface density current $J^r$ for the model illustrated in Fig.\ \ref{fig:my_label5} (planar model with monoenergetic particles;  $\varepsilon_0 = 1.3$ and $\xi_0 = 20$). The exact solution (Eq.\ (\ref{Jrmonoplanar})) is plotted with the solid line. Blue and green dots show results of a Monte Carlo simulation (Eq.\ (\ref{JrMC2Dmono})). The sample of geodesics orbits is the same as in Fig.\ \ref{fig:my_label5}. There are $5 \times 10^4$ orbits: 15953 absorbed trajectories and 34047 scattered ones.}
    \label{fig:my_label6}
\end{figure}

\begin{figure}
    \centering
    \includegraphics[width=\columnwidth]{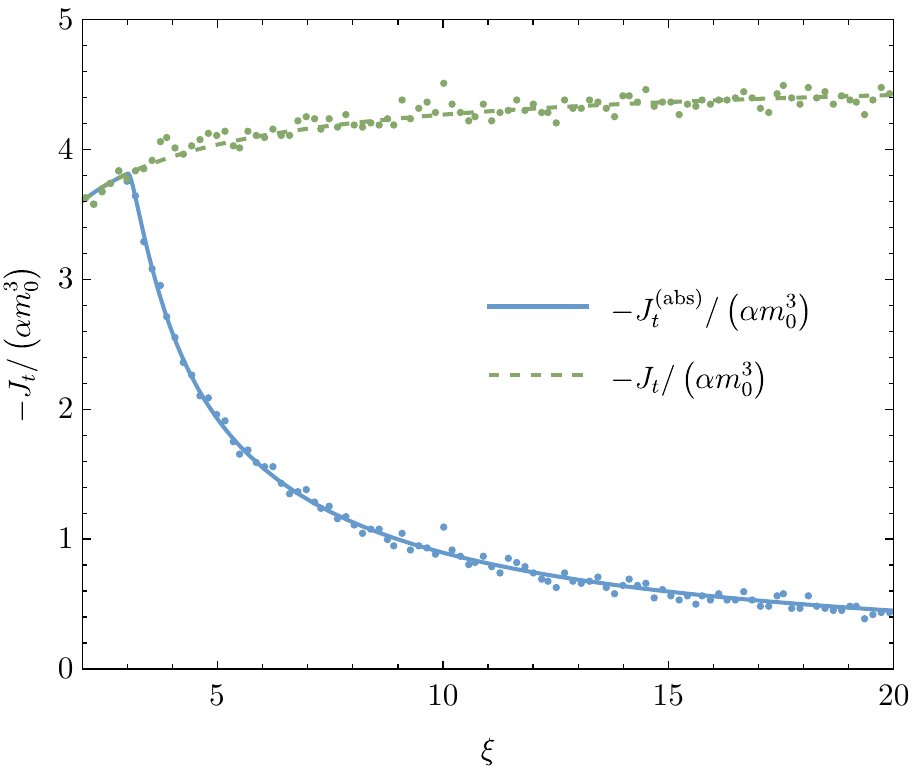}
    \caption{Time components of the particle surface density current $J_t$ for the planar Maxwell-J\"{u}ttner model. In this case $\beta = 1$, $\varepsilon_\mathrm{cutoff} = 10$, and $\xi_0 = 20$. Exact solutions (Eqs.\ (\ref{Jtabsexpplanar}) and (\ref{Jtscatexpplanar})) are plotted with solid and dashed lines. Blue and green dots show results of a Monte Carlo simulation (Eqs.\ \ref{Jtexp2D}). There are $2 \times 10^6$ sets $(\varepsilon_i, \lambda_i, \varphi_{0,i}, \epsilon_{\lambda,i}, y_i)$, giving $10356$ absorbed and $27056$ scattered orbits.}
    \label{fig:my_label5b}
\end{figure}

\begin{figure}
    \centering
    \includegraphics[width=\columnwidth]{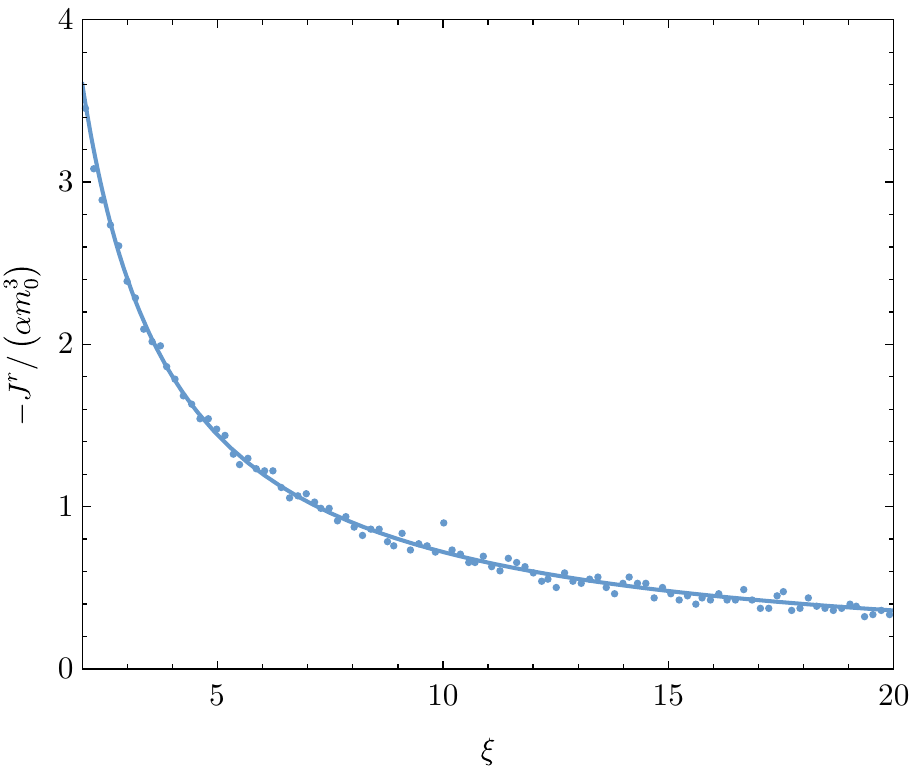}
    \caption{The radial component of the particle surface density current $J_r$ for the planar accretion model with a Maxwell-J\"{u}ttner asymptotic distribution. In this case $\beta = 1$, $\xi_0 = 20$, $\varepsilon_\mathrm{cutoff} = 10$. The exact solution (Eq.\ (\ref{Jrexpplanar})) is plotted with a solid line. Blue dots show results of a Monte Carlo simulation (Eq.\ (\ref{Jrexp2D})). The sample of geodesics orbits is the same as in Fig.\ \ref{fig:my_label5b}. There are $2 \times 10^6$ sets $(\varepsilon_i, \lambda_i, \varphi_{0,i}, \epsilon_{\lambda,i}, y_i)$, giving $10356$ absorbed and $27056$ scattered orbits.}
    \label{fig:my_label6b}
\end{figure}

\subsubsection{Averages}

We construct our Monte Carlo simulations of stationary flows by selecting a set of geodesic trajectories and counting their intersections with suitably chosen surfaces, assuming appropriate weights.

As introductory examples, consider planar problems described in Sec.\ \ref{sec:planar}. We aim at a Monte Carlo simulation that would allow us to compute the components of $J_\mu$ (say). This poses several problems. The first one is related to the choice of a convenient parametrization of geodesics. The second is to select geodesics assuming the correct probability distribution expressed in terms of geodesic parameters. Finally, one has to design a method of computing suitable averages over momenta. We will start the discussion with this last problem.

The essence of our Monte Carlo simulations is an approximation of a continuous system by a discrete one, given be the distribution function
\begin{eqnarray}
\mathcal F (x^\mu, p_\nu) & = & \int \sum_{i = 1}^N \delta^{(4)} \left( x^\mu - x^\mu_{(i)}(\tau) \right) \nonumber \\
&& \times \delta^{(4)} \left( p_\nu - p_\nu^{(i)}(\tau) \right) d\tau,
\label{dfdiscrete}
\end{eqnarray}
where $N$ denotes the number of particles with trajectories described by coordinates $x^\mu_{(i)}(\tau)$ and momenta $p_\nu^{(i)}(\tau)$ (cf.\ \cite{Groot}, p.\ 14, Eq.\ (A6)). Our convention regarding the parameter $
\tau$ is that of Eq.\ (\ref{hamiltonseqs}), and hence Eq.\ (\ref{dfdiscrete}) differs by the factor $1/m$ with respect to the formula in \cite{Groot}. The particle current density $\mathcal J_\mu$ corresponding to this distribution can be computed as
\begin{eqnarray}
    \mathcal J_\mu(x) & = & \int \mathcal F (x,p) p_\mu \sqrt{- \mathrm{det} \, g^{\alpha \beta}(x)} dp_0 \dots dp_3 \nonumber \\
    & = & \int  \sum_{i = 1}^N  \delta^{(4)} \left( x^\alpha - x^\alpha_{(i)}(\tau) \right)  \nonumber \\
    && \times p_\mu^{(i)} (\tau) \sqrt{- \mathrm{det} \, g^{\alpha \beta}(x)} d \tau.
\end{eqnarray}

The above expression is compatible with Eq.\ (\ref{NJ}), which can be seen as follows. Let $S$ denote a hypersurface of constant time, and let the spacetime metric $g$ in the vicinity of $S$ be expressed in the standard $3+1$ form
\begin{equation}
g = (-\alpha^2 + \beta_i \beta^i) dt^2 + 2 \beta_i dt dx^i + \gamma_{ij}dx^i dx^j, \end{equation}
where $\alpha$ is the lapse, $\beta^i$ are the components of the shift vector, and $\gamma$ denotes the metric induced on $S$. Then,
\begin{eqnarray}
\mathcal N[S] & = & - \int_S \mathcal J_\mu s^\mu \eta_S \nonumber \\
& = & - \int_S \eta_S \int d \tau \sum_{i=1}^N \delta^{(4)} \left( x^\mu - x^\mu_{(i)}(\tau) \right) \nonumber \\
&& \times \sqrt{- \mathrm{det} \, g^{\delta \kappa}(x)} p^\nu_{(i)}(\tau) s_\nu. 
\end{eqnarray}
Here $s_\nu = (-\alpha,0,0,0)$, and $\eta_S = \sqrt{\mathrm{det} \, \gamma_{ij}} d^3x$. Thus,
\begin{eqnarray}
\mathcal N[S] & = & \int_S d^3 x \int d \tau \sum_{i=1}^N \delta^{(4)} \left( x^\mu - x^\mu_{(i)}(\tau) \right) p^t_{(i)}(\tau) \nonumber \\
& = & \sum_{i=1}^N \int d \tau \delta \left( t - t_{(i)}(\tau) \right) \frac{d t_{(i)}}{d \tau} = N,
\end{eqnarray}
where we have used the fact that $\sqrt{- \mathrm{det} \, g_{\mu \nu}} = \alpha \sqrt{\mathrm{det} \, \gamma_{ij}}$.

We will estimate the value of $\mathcal J_\mu$ at a point $x$, by selecting a (small) hypersurface region $\Sigma$ (a cell), such that $x \in \Sigma$, and computing
\begin{equation}
\langle \mathcal J_\mu \rangle = \frac{\int_\Sigma \mathcal J_\mu \eta_\Sigma}{\int_\Sigma \eta_\Sigma}.
\end{equation}
For planar models, we might be interested in the surface density value
\begin{equation}
\langle J_\mu \rangle = \frac{\int_\Sigma J_\mu \eta_\Sigma}{\int_\Sigma \eta_\Sigma} \end{equation}
as well.

Perhaps the most natural method would be to select a region within a hypersurface of constant time, discretize this region into cells, and average over trajectories passing through a given cell. The trouble with this approach is that it is not well suited to searching for stationary solutions. Instead, we can select a timelike hypersurface foliated by the orbits of a stationary Killing field, and count the intersections of the geodesics with such a surface, assuming additional weights which we derive below.

For a planar, stationary accretion flow in the Schwarzschild spacetime, we take a segment
\begin{equation}
S = \{ (r,\theta,\varphi) \colon r_1 \le r \le r_2, \, \theta = \pi/2, \, \varphi = \varphi_0 \}
\end{equation}
and a surface
\begin{eqnarray}
\Sigma & = & \{ (t, r,\theta,\varphi) \colon  t_1 \le t \le t_2, \, r_1 \le r \le r_2, \nonumber \\
&& \theta = \pi/2, \, \varphi = \varphi_0 \}.
\end{eqnarray}
Let $\Phi_\tau(x_0^i)$ denote the orbit of the timelike Killing vector field $\xi^\mu = (1,0,0,0)$, passing through $x_0^i$ at $\tau = 0$, i.e., $\Phi_0(x_0^i) = x_0^i$. Then $\Sigma$ can be expressed as the image
\begin{equation}
\Sigma = \Phi_{[t_1,t_2]}(S).
\end{equation}

For the planar model we have
\begin{eqnarray}
\mathcal J_\mu (x) & = &  \int \sum_{i = 1}^N \delta(\theta - \pi/2)\delta^{(3)} \left( x^\alpha - x^\alpha_{(i)}(\tau) \right) \nonumber \\
&& \times p_\mu^{(i)}(\tau) \sqrt{- \mathrm{det} \, g^{\alpha \beta}(x)} d \tau,
\end{eqnarray}
where
\begin{eqnarray}
\delta^{(3)} \left( x^\alpha - x^\alpha_{(i)}(\tau) \right) & = & \delta \left( t - t_{(i)} (\tau) \right) \delta \left( r - r_{(i)}(\tau) \right)  \nonumber \\
&& \times \delta \left( \varphi_0 - \varphi_{(i)}(\tau) \right).
\end{eqnarray}
Since, at the equatorial plane $\delta(\theta - \pi/2) = r \delta(z)$ and $\mathcal J_\mu = \delta(z) J_\mu$, we get
\begin{eqnarray}
J_\mu (x) & = &  \int \sum_{i = 1}^N r \delta^{(3)} \left( x^\alpha - x^\alpha_{(i)}(\tau) \right) \nonumber \\
&& \times p_\mu^{(i)}(\tau) \sqrt{- \mathrm{det} \, g^{\alpha \beta}(x)} d \tau,
\end{eqnarray}
where $\sqrt{- \mathrm{det} \, g^{\alpha \beta}(x)} = \left( r^2 \sin \theta \right)^{-1} = 1/r^2$. The volume element induced on $\Sigma$ reads $\eta_\Sigma = dt dr$. A direct calculation gives
\begin{eqnarray}
\int_\Sigma J_\mu \eta_\Sigma & = & \int_{t_1}^{t_2} dt \int_{r_1}^{r_2} dr \int d \tau \sum_{i = 1}^N \delta^{(3)} \left( x^\alpha - x^\alpha_{(i)}(\tau) \right) \nonumber \\
&& \times \frac{p_\mu^{(i)}(\tau)}{r} \nonumber \\
& = & \int d \tau  \sum_{i = 1}^{N} \delta \left( \varphi_0 - \varphi_{(i)}(\tau) \right) \frac{p_\mu^{(i)}(\tau)}{r_{(i)}(\tau)},
\end{eqnarray}
where in the last sum we only take into account trajectories that intersect $\Sigma$. The integral with respect to $\tau$ can be computed by writing
\begin{equation}
\delta \left( \varphi_0 - \varphi_{(i)}(\tau) \right) = \sum_{k} \frac{\delta (\tau - \tau_k)}{ \left| \left. \frac{d \varphi_{(i)}}{d \tau} \right|_{\tau = \tau_k} \right|},
\end{equation}
where $\varphi_{(i)}(\tau_k) = \varphi_0$. The above sum runs over all intersections of the $i$-th trajectory with $\Sigma$. Note that
\begin{equation}
\frac{d \varphi_{(i)}}{d \tau} = g^{\varphi \varphi} p_\varphi^{(i)} = \frac{1}{r_{(i)}^2} p_\varphi^{(i)}.
\end{equation}
Thus, we get finally,
\begin{equation}
\int_\Sigma J_\mu \eta_\Sigma = \sum_{i = 1}^{N_\mathrm{int}} \frac{p_\mu^{(i)} r_{(i)}}{l_{(i)}},
\end{equation}
where the $N_\mathrm{int}$ is the number of \textit{all intersections} of trajectories with the surface $\Sigma$. The particle current surface density can be now approximated as
\begin{eqnarray}
\langle J_\mu \rangle & = & \frac{1}{(t_2 -t_1) (r_2 - r_1)} \sum_{i = 1}^{N_\mathrm{int}} \frac{p_\mu^{(i)} r_{(i)}} {l_{(i)}} \nonumber \\
& = & \frac{1}{M m (t_2 -t_1) (\xi_2 - \xi_1)} \sum_{i = 1}^{N_\mathrm{int}} \frac{p_\mu^{(i)} \xi_{(i)}} {\lambda_{(i)}}.
\label{Jmuplanar}
\end{eqnarray}
For stationary problems, the result should be independent of the choice of $t_1$ and $t_2$ in a sense that the number of trajectories that intersect $\Sigma$ should be proportional to the length $t_2 - t_1$, if the latter is sufficiently large. In practice, we omit the factor $t_2 - t_1$ and normalize the results by the number of trajectories taken into account. Moreover, instead of considering full orbits in the four dimensional spacetime, it is sufficient to work with projections of trajectories onto surfaces of constant time $t$.

\subsubsection{Selecting trajectories}

The setup described above brings us immediately to the problem of selecting the appropriate distribution of geodesics. Let us start the discussion with a homogeneous distribution of a gas within a plane in the Minkowski spacetime, introduced in Sec.\ \ref{sec:flatminkowski}. In principle, we are interested in an infinite distribution, but for practical reasons, a Monte Carlo simulation has to be restricted to a compact region in space, say a disk of a dimensionless radius $\xi_0$. The problem we are facing turns out to be a variation of a the classic Bertrand problem (or Bertrand paradox), known from the theory of geometrical probability \cite{Bertrand}. Bertrand's problem is usually formulated as follows. On a fixed circle, one randomly selects a chord. What is the probability that the length of this chord is larger than the side of an equilateral triangle inscribed in this circle? It is then shown that different methods of selecting the chord ``at random'' lead to different answers (different probabilities). One of the methods of selecting chords is to demand that they belong to straight lines uniformly distributed in a plane. Such a distribution has been studied, e.g., by Kendall and Moran in \cite{kendall}. In terms of Cartesian coordinates $(x,y)$ straight lines can be parametrized by
\begin{equation}
ux + vy + 1 = 0.
\end{equation}
It can be shown that the probability distribution
\begin{equation}
F(u,v) du dv = (u^2 + v^2)^{-\frac{3}{2}} du dv
\end{equation}
is invariant under the group of Euclidean symmetries on the plane---rotations and translations \cite{kendall}. The trouble with parameters $u$ and $v$ is that they are not well suited to a description of chords within a given circle. Instead a better description can be provided by parametrizing the chords (or equivalently straight lines in the plane) by polar coordinates $(p,\vartheta)$ of the point on the chord with a smallest distance to the center of the circle. In terms of $(p,\vartheta)$, one has
\begin{equation}
\label{poincare1}
(u^2 + v^2)^{-\frac{3}{2}} du dv = dp d\vartheta,
\end{equation}
and given that $du dv = p  dp  d\vartheta$, we get
\begin{equation}
\label{poincare2}
F(p,\vartheta) = p^{-1}.
\end{equation}
In the context of Bertand's paradox this distribution has already been proposed by Poincar\'{e} in 1912 \cite{Poincare}.

While in the original formulation of Bertrnad's paradox choosing a solution may be a matter of preference, Poincar\'{e}'s solution happens to be well suited to our problem at hand, as the restriction to a given circle is only technical---we are in fact interested in a description of a gas of particles with trajectories distributed uniformly in a plane. For monoenergetic particles of energy $\varepsilon_0$, the parametrization in terms of the coordinate $p$ is equivalent to the one with the total angular momentum $\lambda$, as $\lambda$ and
\begin{equation}
\label{pxi}
p = M \xi =\frac{ M \lambda}{\sqrt{\varepsilon_0^2 - 1}}
\end{equation}
are linearly related. Let us now return to the angular momentum distribution appearing in Eqs.\ (\ref{Jtmink1}) and (\ref{Jtmink3}). Using Eq.\ (\ref{pxi}), we can express $J_t$ given by Eqs.\ (\ref{Jtmink1}) and (\ref{Jtmink3}) as
\begin{equation}
    J_t = - \frac{\alpha m_0^3 \varepsilon_0}{M \xi} \int_0^{M \xi} dp \int_0^{2 \pi} d \vartheta.
\end{equation}

To conclude: in a Monte Carlo simulation of uniformly distributed monoenergetic particles confined to a plane in a flat Minkowski spacetime, we would select trajectories distributed uniformly in $\lambda$. An example of such a distribution of straight lines in a plane is shown in Fig.\ \ref{fig:my_label1}. Since the total angular momentum $\lambda$ is a constant of motion and our models of accretion in the Schwarzshcild spacetime assume a uniform planar asymptotic distribution of particles, a uniform distribution of $\lambda$ has to be assumed for these models as well.

\subsubsection{Implementation for the planar accretion problem}

Unbound trajectories confined to a plane in the Schwarzschild spacetime can be parametrized by the energy $\varepsilon$, angular momentum $\lambda$, the location of the intersection of the orbit with the circle of the given radius $\xi_0$ (outer boundary of the simulation region), and the sign $\epsilon_\lambda$ distinguishing between the clockwise or anticlockwise direction of motion. If $\lambda = |\lambda_z|$ (the motion occurs in the equatorial plane), we set $\epsilon_\lambda = \mathrm{sgn} \, \lambda_z$. We compute trajecotries using a formalism based on the Biermann-Weierstrass theorem, introduced in \cite{cieslik_mach_2022}. Its main ingredient is a universal formula for a radius in terms of the so-called true anomaly angle $\psi$ (the polar angle in the orbital plane). It reads
\begin{widetext}
\begin{equation}
\label{xi_psi}
    \xi(\psi) =  \xi_0 + \frac{- \epsilon_r \sqrt{f(\xi_0)} \wp'(\psi) + \frac{1}{2} f'(\xi_0 ) \left[ \wp(\psi) - \frac{1}{24}f''(\xi_0 )\right] + \frac{1}{24} f(\xi_0 ) f'''(\xi_0 )  }{2 \left[ \wp(\psi) - \frac{1}{24} f''(\xi_0 ) \right]^2 - \frac{1}{48} f(\xi_0 ) f^{(4)}(\xi_0 ) },
\end{equation}
\end{widetext}
where 
\begin{equation}
\label{f_general}
    f(\xi) = a_0 \xi^4 + 4a_1 \xi^3 +6a_2\xi^2 + 4a_3\xi,
\end{equation}
and
\begin{equation}
\label{fcoeffs}
    a_0 = \frac{\varepsilon^2 -1}{\lambda^2}, \quad 4a_1 = \frac{2}{\lambda^2}, \quad 6a_2 = - 1, \quad 4a_3 = 2.
\end{equation}
The Weierstrass $\wp$ function has to be computed with the following Weierstrass invariants
\begin{subequations}
 \label{invariants_phys}
\begin{eqnarray}   
        g_2 & = & \frac{1}{12} - \frac{1}{\lambda^2}, \\
        g_3 & = & \frac{1}{6^3} - \frac{1}{12 \lambda^2} - \frac{\varepsilon^2 - 1}{4 \lambda^2},
\end{eqnarray}
\end{subequations}
i.e., $\wp(\psi) = \wp(\psi; g_2, g_3)$, and the same applies to the derivative $\wp'(\psi)$. Here the sign $\epsilon_r = \pm 1$ corresponds to the radial direction of motion at the initial location $\xi_0$. Taking ingoing trajectories at $\xi_0$, we set $\epsilon_r = -1$.

As another element, we need formulas for the allowed range $R_\psi$ of the true anomaly $\psi$. It is computed differently for absorbed and scattered orbits. We define the following function
\begin{equation}
    X(\xi) = \int_\xi^\infty \frac{d \xi^\prime}{\sqrt{ f(\xi^\prime)}}.
\end{equation}

For generic unbound scattered orbits $X(\xi)$ can be expressed as
\begin{widetext}
\begin{equation}
    X(\xi) = \frac{1}{\sqrt{y_3 - y_1}} \left[ \tilde F \left( \arccos \sqrt{ \frac{y_2 + \frac{1}{12} - \frac{1}{2\xi}}{y_2 - y_1}} , k \right) - \tilde F \left( \arccos \sqrt{\frac{y_2 + \frac{1}{12}}{y_2 - y_1}}, k \right) \right],
\end{equation}
\end{widetext}
where $y_1$, $y_2$, and $y_3$ denote the roots of the polynomial $4y^3 - g_2 y - g_3$, satisfying $y_1 < y_2 < y_3$,
\begin{equation}
k^2 = \frac{y_2 - y_1}{y_3 - y_1},
\end{equation}
and $\tilde F(\phi,k)$ is the Legendre elliptic integral defined by
\begin{equation}
    \tilde F(\phi,k) = \int_0^\phi \frac{d \chi}{\sqrt{1 - k^2 \sin^2 \chi}}, \quad -\frac{\pi}{2} < \phi < \frac{\pi}{2}.
\end{equation}

For unbound absorbed orbits an explicitly real formula for $X(\xi_0)$ is different. Let $y_1$ denote a real zero of the polynomial $4y^3 - g_2 y - g_3 = 4 (y - y_1)(y^2 + py + q)$, where $p^2 - 4 q < 0$ and thus $y^2 +p y + q > 0$. We define $\mu = \sqrt{y_1^2 + p y_1 + q}$ and
\begin{equation}
\label{k2bis}
    k^2 = \frac{1}{2} \left( 1 - \frac{y_1 + p/2}{\mu} \right).
\end{equation}
The expression for $X(\xi)$ reads
\begin{widetext}
\begin{equation}
\label{xbis}
    X(\xi) = \frac{1}{2 \sqrt{\mu}} \left[ \tilde F\left(2 \arctan \sqrt{\frac{- \frac{1}{12} + \frac{1}{2 \xi} - y_1}{\mu}} , k \right) - \tilde F \left( 2 \arctan \sqrt{\frac{-\frac{1}{12} - y_1}{\mu}} , k \right) \right].
\end{equation}
\end{widetext}
The calculation of the allowed range of $\psi$ is now quite simple. For unbound scattered orbits one has
\begin{equation}
R_\psi = 2 [X(\xi_\mathrm{peri}) - X(\xi_0)],
\end{equation}
where $\xi_\mathrm{peri}$ denotes the pericenter radius. For unbound absorbed orbits we set
\begin{equation}
R_\psi = X(2) - X(\xi_0),
\end{equation}
since $\xi = 2$ corresponds to the black hole horizon.

Note that the range $R_\psi$ of the angle $\psi$ can be larger than $2 \pi$, in which case the orbit circles around the black hole. Thus, the last ``technical'' element in our description is a simple function that gives the number of intersections of a given geodesic trajectory with a radius of constant polar angle $\varphi$. To define such a function one only needs to know the range $R_\psi$, the polar coordinate $\varphi_0$ corresponding to the initial radius $\xi_0$, the sign $\epsilon_\lambda$, and the polar coordinate of the radius $\varphi$ for which one counts the intersections with the given geodesic. This idea is illustrated in Fig.\ \ref{fig:my_label2}. In this case a sample absorbed geodesic is depicted with a blue line. Radial segments which are intersected twice are marked with solid gray lines. Radial segments with three intersections are marked with dashed lines. A sample Wolfram Mathematica \cite{Mathematica} code for such a function looks as follows (here $\mathtt{phi} = \varphi$, $\mathtt{phi0} = \varphi_0$, $\mathtt{range} = R_\psi$, $\mathtt{dir} = \epsilon_\lambda$):
\begin{widetext}
\begin{verbatim}
NumberOfIntersections[phi_, phi0_, range_, dir_] := If[dir == 1,
  If[phi > phi0, 
   If[Mod[range, 2 Pi] > phi - phi0, Quotient[range, 2 Pi] + 1, Quotient[range, 2 Pi]], 
   If[Mod[range + phi0, 2 Pi] > phi, Quotient[range + phi0, 2 Pi],
   Quotient[range + phi0, 2 Pi] - 1]],
  If[phi < phi0, 
   If[Mod[range, 2 Pi] > phi0 - phi, Quotient[range, 2 Pi] + 1, Quotient[range, 2 Pi]], 
   If[Mod[phi0 - range, 2 Pi] < phi, Quotient[range - phi0, 2 Pi] + 1,
   Quotient[range - phi0, 2 Pi]]]
  ]
\end{verbatim}
\end{widetext}

Given all these elements, we select trajectories with a given energy $\varepsilon_0$ and uniformly distributed values of $\varphi_0 \in [0,2 \pi)$, $\lambda \in [0, \lambda_\mathrm{max}(\varepsilon_0,\xi_0)]$, and the sign $\epsilon_\lambda = \pm 1$. For clarity, these parameters can be divided into two classes: those corresponding to absorbed trajectories with $\lambda < \lambda_c(\varepsilon)$ and scattered orbits with $\lambda > \lambda_c(\varepsilon)$. A plot of such trajectories, projected at a hypersurface of constant time, is given in Fig.\ \ref{fig:my_label3}.

According to Eq.\ (\ref{Jmuplanar}), the components of $J_\mu$ can be computed by counting the number of intersections per length interval at a given radius ($\varphi = \mathrm{const}$), with appropriate weights. Such intersections are illustrated in Fig.\ \ref{fig:my_label4}. Let the set of indices $I_\mathrm{abs}(\xi_1,\xi_2)$ number the intersections of absorbed trajectories in the segment $\xi_1 \le \xi \le \xi_2$ of a radius of constant $\varphi$. Note that a given trajectory can contribute more than once to the set of intersections $I_\mathrm{abs}(\xi_1,\xi_2)$---this can happen for spiraling orbits, similar to the one shown in Fig.\ \ref{fig:my_label2}. An analogous set of indices referring to scattered orbits will be denoted by $I_\mathrm{scat}(\xi_1,\xi_2)$. The Monte Carlo estimators of components $J_t^\mathrm{(abs)}$ and $J_t^\mathrm{(scat)}$ at $(\xi = (\xi_1 + \xi_2)/2, \varphi)$ can be computed as
\begin{subequations}
\label{JtMC2Dmono}
\begin{eqnarray}
\langle J_t^\mathrm{(abs)} \rangle & = &- \frac{4 \pi \alpha m_0^3 \lambda_c(\varepsilon_0)}{N_\mathrm{abs} (\xi_2 - \xi_1)} \sum_{i \in I_\mathrm{abs}(\xi_1,\xi_2)} \frac{\varepsilon_0 \xi_{(i)}
}{\lambda_{(i)}}, \\
\langle J_t^\mathrm{(scat)} \rangle & = & - \frac{4 \pi \alpha m_0^3 [ \lambda_\mathrm{max}(\varepsilon_0, \xi_0) -\lambda_c(\varepsilon_0) ]}{N_\mathrm{scat} (\xi_2 - \xi_1)} \nonumber \\
&& \times \sum_{i \in I_\mathrm{scat}(\xi_1,\xi_2)} \frac{\varepsilon_0 \xi_{(i)}}{\lambda_{(i)}},
\end{eqnarray}
\end{subequations}
where $N_\mathrm{abs}$ and $N_\mathrm{scat}$ denote the number of absorbed and scattered orbits, respectively. The estimator for $J^r$ can be computed as
\begin{eqnarray}
\langle J^r \rangle & = & - \frac{4 \pi \alpha m_0^3 \lambda_c(\varepsilon_0)}{N_\mathrm{abs} (\xi_2 - \xi_1)} \nonumber \\
&& \times \sum_{i \in I_\mathrm{abs}(\xi_1,\xi_2)} \frac{\xi_{(i)}}{\lambda_{(i)}} \sqrt{\varepsilon_0^2 - U_{\lambda_{(i)}}\left(\xi_{(i)}\right)}.
\label{JrMC2Dmono}
\end{eqnarray}
We discretize the radial direction according to: $\xi_j < \xi < \xi_{j+1}$, $j = 1, \dots, N_\xi - 1$, with $\xi_1 = 2$ and $\xi_{N_\xi} = \xi_0$ (the radius of the outer boundary). Thus, strictly speaking, the values $\xi_1$ and $\xi_2$ appearing in all our formulas should be replaced by $\xi_1 \to \xi_j$, $\xi_2 \to \xi_{j+1}$. We will, however, keep the notation with $\xi_1$ and $\xi_2$ referring to cell boundaries, as it is probably more transparent. A comparison of results obtained according to Eqs.\ (\ref{JtMC2Dmono}) and (\ref{JrMC2Dmono}) with exact expressions for $J_t$ and $J^r$ is shown in Figs.\ \ref{fig:my_label5} and \ref{fig:my_label6}.

Apart from the number of trajectories, the resolution of the method is controlled by the number of radii of constant $\varphi$ and the size $\xi_{j+1} - \xi_j$ of radial segments $[\xi_j,\xi_{j+1}]$. In the example shown in Figs.\ \ref{fig:my_label5} and \ref{fig:my_label6} there are 120 radii of constant $\varphi$, distributed within the full angle, and 100 equidistant radial segments $[\xi_j,\xi_{j+1}]$ in the range $2 < \xi \le \xi_0 = 20$ at each radius $\varphi = \mathrm{const}$. The values of $\langle J_\mu \rangle$ can be computed separately for each radius of constant $\varphi$; in our examples we additionally average the results over all radii of constant $\varphi$.

In a similar fashion one can compute Monte Carlo estimators of the particle current density in the case with the Maxwell-J\"{u}ttner asymptotic distribution. It is convenient to select trajectories directly from the Maxwell-J\"{u}ttner distribution. Again, von Neumann's rejection method can be used. We repeat the procedure described in Sec.\ \ref{sec:montecarlointegration}, but this time four parameters $(\varepsilon_i,\lambda_i,\varphi_{0,i},\epsilon_{\lambda,i},y_i)$ are selected in each iteration step, and uniform distributions are assumed for $\varphi_{0,i} \in [0,2\pi)$, and $\epsilon_{\lambda,i} = \pm 1$.

Monte Carlo estimators of the particle current surface density in the asymptotically Maxwell-J\"{u}ttner model can be written as
\begin{subequations}
\label{Jtexp2D}
\begin{eqnarray}
\langle J_t^\mathrm{(abs)} \rangle & = &- \frac{4 \pi \alpha m_0^3 V_\mathrm{abs}}{N_\mathrm{abs} (\xi_2 - \xi_1)}  \sum_{i \in I_\mathrm{abs}(\xi_1,\xi_2)} \frac{\varepsilon_{(i)} \xi_{(i)}}{\lambda_{(i)}}, \\
\langle J_t^\mathrm{(scat)} \rangle & = & - \frac{4 \pi \alpha m_0^3 V_\mathrm{scat}}{N_\mathrm{scat} (\xi_2 - \xi_1)} \sum_{i \in I_\mathrm{scat}(\xi_1,\xi_2)} \frac{\varepsilon_{(i)} \xi_{(i)}}{\lambda_{(i)}},
\end{eqnarray}
\end{subequations}
where $V_\mathrm{abs}$ and $V_\mathrm{scat}$ are given by Eqs.\ (\ref{volumes}). The estimator for $J^r$ can be computed as
\begin{eqnarray}
\langle J^r \rangle & = & - \frac{4 \pi \alpha m_0^3 V_\mathrm{abs}}{N_\mathrm{abs} (\xi_2 - \xi_1)} \nonumber \\
&& \times \sum_{i \in I_\mathrm{abs}(\xi_1,\xi_2)} \frac{\xi_{(i)}}{\lambda_{(i)}} \sqrt{\varepsilon_{(i)}^2 - U_{\lambda_{(i)}}\left(\xi_{(i)}\right)}.
\label{Jrexp2D}
\end{eqnarray}
A comparison of these estimators and exact solutions given by Eqs.\ (\ref{Jexpplanar}) is shown in Fig.\ \ref{fig:my_label5b} and \ref{fig:my_label6b}.

\section{Spherically symmetric solutions}
\label{sec:spherical}

\subsection{Spherically symmetric solutions in the Schwarzschild background}

We now turn to spherically symmetric solutions. The calculations given in this section are similar to those for planar systems, and thus we will mostly only summarize the results. Spherically symmetric models describing stationary accretion of collisionless gas in the Schwarzschild spacetime were derived in \cite{Olivier1,Olivier2}, in particular assuming the Maxwell-J\"{u}ttner distribution at infinity. In the following two subsections we give expressions for the particle current density in the spherically symmetric stationary accretion model with monoenergetic particles and recall expressions derived already in \cite{Olivier1,Olivier2} for the spherical Maxwell-J\"{u}ttner model. 

\subsubsection{Monoenergetic particles}

For the gas of monoenergetic particles, the one-particle distribution function reads, in our standard setup,
\begin{eqnarray}
\mathcal F(x, p) & = & \alpha m_0 \delta\left(\sqrt{- p_\mu p^\mu} - m_0\right) \delta(p_t + E_0) \nonumber \\
& = & \alpha \delta\left(\sqrt{- p_\mu p^\mu} - m_0\right)  \delta(\varepsilon - \varepsilon_0),
\end{eqnarray}
where, as before, we only take into account future pointing momenta. The momentum-space volume element can be written as
\begin{equation}
\mathrm{dvol}_x(p) = \frac{m^3 \lambda}{\xi^2 \sqrt{\varepsilon^2 - U_\lambda(\xi)}} d \varepsilon dm d\lambda d\chi,
\end{equation}
where the variable $\chi$ is defined as
\begin{equation}
p_\theta = M m \lambda \cos \chi, \quad p_\varphi = M m \lambda \sin \theta \sin \chi
\end{equation}
(see \cite{Olivier1}).

Taking into account the regions in the phase space available for the motion of absorbed and scattered particles, one can write time components of the particle current density as
\begin{widetext}
\begin{subequations}
\begin{eqnarray}
    \mathcal J_t^\mathrm{(abs)}(\xi) & = & - \frac{2 \pi \alpha m_0^4 \varepsilon_0}{\xi^2} \Theta(\varepsilon_0 - 1) \int_0^{\lambda_c(\varepsilon_0)} \frac{\lambda d \lambda}{\sqrt{\varepsilon_0^2 - U_\lambda (\xi)}}, \\
    \mathcal J_t^\mathrm{(scat)}(\xi) & = & - \frac{4 \pi \alpha m_0^4 \varepsilon_0}{\xi^2} \Theta(\varepsilon_0 - \varepsilon_\mathrm{min}(\xi)) \int_{\lambda_c(\varepsilon_0)}^{\lambda_\mathrm{max}(\varepsilon_0,\xi)} \frac{ \lambda d \lambda}{\sqrt{\varepsilon_0^2 - U_\lambda (\xi)}}.
\end{eqnarray}
\end{subequations}
The integrals with respect to momenta can be easily computed, yielding
\begin{subequations}
\label{Jt3DmonoEq}
\begin{eqnarray}
\mathcal J_t^\mathrm{(abs)}(\xi) & = & - \frac{2 \pi \alpha m_0^4 \varepsilon_0}{\xi^2} \Theta(\varepsilon_0 - 1) \frac{\lambda_c^2(\varepsilon_0)}{s_{\lambda_c}(\varepsilon_0,\xi) + s_0(\varepsilon_0,\xi)}, \\
\mathcal J_t^\mathrm{(scat)}(\xi) & = & - \frac{4 \pi \alpha m_0^4 \varepsilon_0}{\xi^2} \Theta(\varepsilon_0 - \varepsilon_\mathrm{min}(\xi)) \frac{\lambda^2_\mathrm{max}(\varepsilon_0,\xi) - \lambda_c(\varepsilon_0)^2}{s_{\lambda_c}(\varepsilon_0,\xi) + s_{\lambda_\mathrm{max}}(\varepsilon_0,\xi)},
\end{eqnarray}
\end{subequations}
\end{widetext}
where
\begin{equation} s_\lambda(\varepsilon,\xi) = \sqrt{\varepsilon^2 - U_\lambda(\xi)}.
\end{equation}
Similarly, for $\mathcal J^r$, we get
\begin{eqnarray}
    \mathcal J^r(\xi) & = & - \frac{2 \pi \alpha m_0^4}{\xi^2} \Theta(\varepsilon_0 - 1) \int_0^{\lambda_c(\varepsilon_0)} \lambda d \lambda \nonumber \\
    & = & - \frac{\pi \alpha m_0^4}{\xi^2} \Theta(\varepsilon_0 - 1) \lambda_c(\varepsilon_0)^2.
    \label{Jr3DmonoEq}
\end{eqnarray}

\subsubsection{Asymptotic Maxwell-J\"{u}ttner distribution}

Assuming the gas obeying the Maxwell-J\"{u}ttner distribution asymptotically, we get the one-particle distribution function in the form
\begin{eqnarray}
\mathcal F(x, p) & = & \alpha \delta\left(\sqrt{- p_\mu p^\mu} - m_0\right) \exp \left( \frac{\beta}{m} k^\mu p_\mu \right) \nonumber \\
& = & \alpha \delta\left(\sqrt{- p_\mu p^\mu} - m_0\right) \exp \left( - \beta \varepsilon \right).
 \end{eqnarray}
 This gives the following expressions for $\mathcal J_t$ and $\mathcal J^r$.
\begin{widetext}
\begin{subequations}
\begin{eqnarray}
\mathcal J_t^{\mathrm{(abs)}}(\xi) & = & - \frac{2 \pi \alpha m_0^4}{\xi^2} \int_1^\infty d\varepsilon e^{- \beta \varepsilon} \varepsilon \int_0^{\lambda_c(\varepsilon)} d \lambda \frac{\lambda}{\sqrt{\varepsilon^2 - U_\lambda(\xi)}} = - \frac{2 \pi \alpha m_0^4}{\xi^2} \int_1^\infty d\varepsilon e^{- \beta \varepsilon} \varepsilon \frac{\lambda_c(\varepsilon)^2}{s_{\lambda_c}(\varepsilon,\xi) + s_0(\varepsilon,\xi)}, \label{Jt3DexpEqabs} \\
\mathcal J_t^{\mathrm{(scat)}}(\xi) & = & - \frac{4 \pi \alpha m_0^4}{\xi^2} \int_{\varepsilon_\mathrm{min}(\xi)}^\infty d\varepsilon e^{- \beta \varepsilon} \varepsilon \int_{\lambda_c(\varepsilon)}^{\lambda_\mathrm{max}(\varepsilon,\xi)} d \lambda \frac{\lambda}{\sqrt{\varepsilon^2 - U_\lambda (\xi) }} \nonumber \\
& = & - \frac{4 \pi \alpha m_0^4}{\xi^2} \int_{\varepsilon_\mathrm{min}(\xi)}^\infty d\varepsilon e^{- \beta \varepsilon} \varepsilon \frac{\lambda_\mathrm{max}(\varepsilon,\xi)^2 - \lambda_c(\varepsilon)^2}{s_{\lambda_c} (\varepsilon,\xi) + s_{\lambda_\mathrm{max}}(\varepsilon,\xi)}, \label{Jt3DexpEqscat} \\
\mathcal J^r(\xi) & = & - \frac{2 \pi \alpha m_0^4}{\xi^2} \int_1^\infty d \varepsilon e^{- \beta \varepsilon} \int_0^{\lambda_c(\varepsilon)} d \lambda \lambda = - \frac{\pi \alpha m_0^4}{\xi^2} \int_1^\infty d \varepsilon e^{- \beta \varepsilon} \lambda_c(\varepsilon)^2. \label{Jr3DexpEq}
\end{eqnarray}
\end{subequations}
\end{widetext}

\subsection{Monte Carlo simulations of spherically symmetric solutions}

\begin{figure}
    \centering
    \includegraphics[width=\columnwidth]{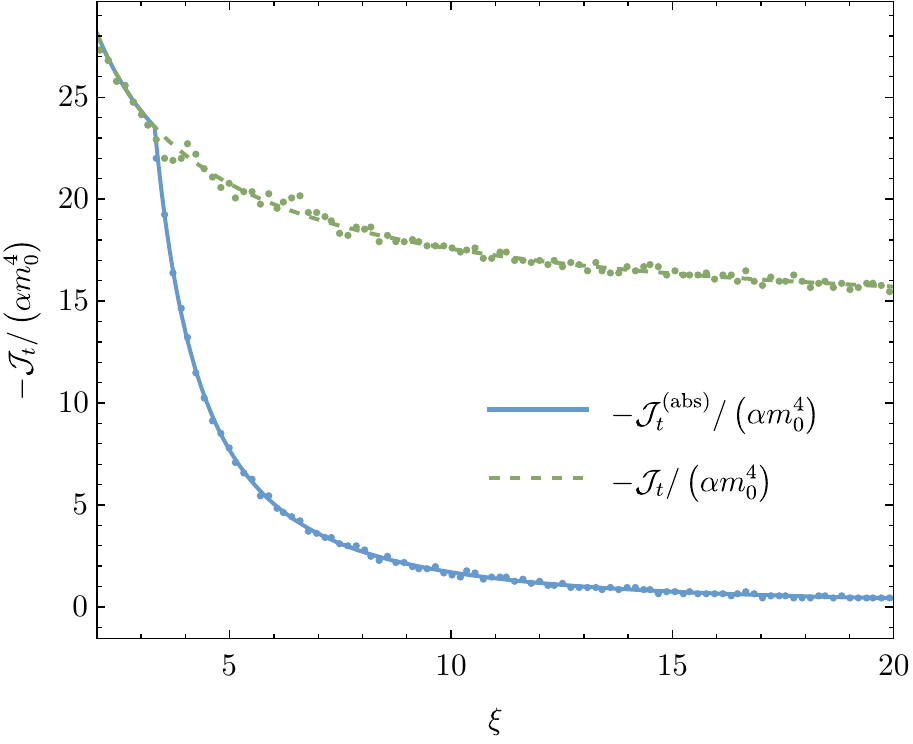}
    \caption{Time components of the particle current density $\mathcal J_t$ for the monoenergetic spherically symmetric model with $\varepsilon_0 = 1.3$ and $\xi_0 = 20$. Exact solutions (Eqs.\ (\ref{Jt3DmonoEq})) are plotted with solid and dashed lines. Blue and green dots show results of a Monte Carlo simulation (Eqs.\ (\ref{Jtmono})). We use von Neumann's rejection method to select appropriate distribution of the total angular momentum. There are $5 \times 10^4$ von Neumann iteration steps giving 2477 absorbed and 22273 scattered trajectories.}
    \label{fig:my_label7}
\end{figure}

\begin{figure}
    \centering
    \includegraphics[width=\columnwidth]{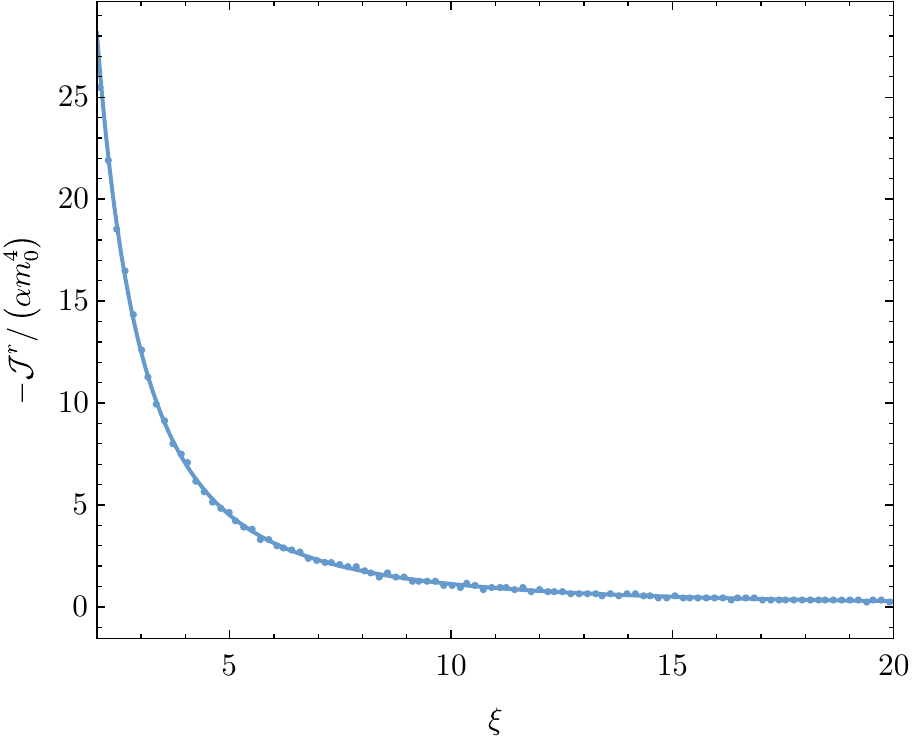}
    \caption{The radial component of the particle current density $\mathcal J^r$ for the monoenergetic spherically symmetric model with $\varepsilon_0 = 1.3$ and $\xi_0 = 20$. The exact solution (Eq.\ (\ref{Jr3DmonoEq})) is plotted with the solid line. Blue dots show results of a Monte Carlo simulation (Eq.\ (\ref{Jrmono})). The sample of geodesics orbits is the same as in Fig.\ \ref{fig:my_label7}. There are 2477 absorbed trajectories.}
    \label{fig:my_label8}
\end{figure}

\begin{figure}
    \centering
    \includegraphics[width=\columnwidth]{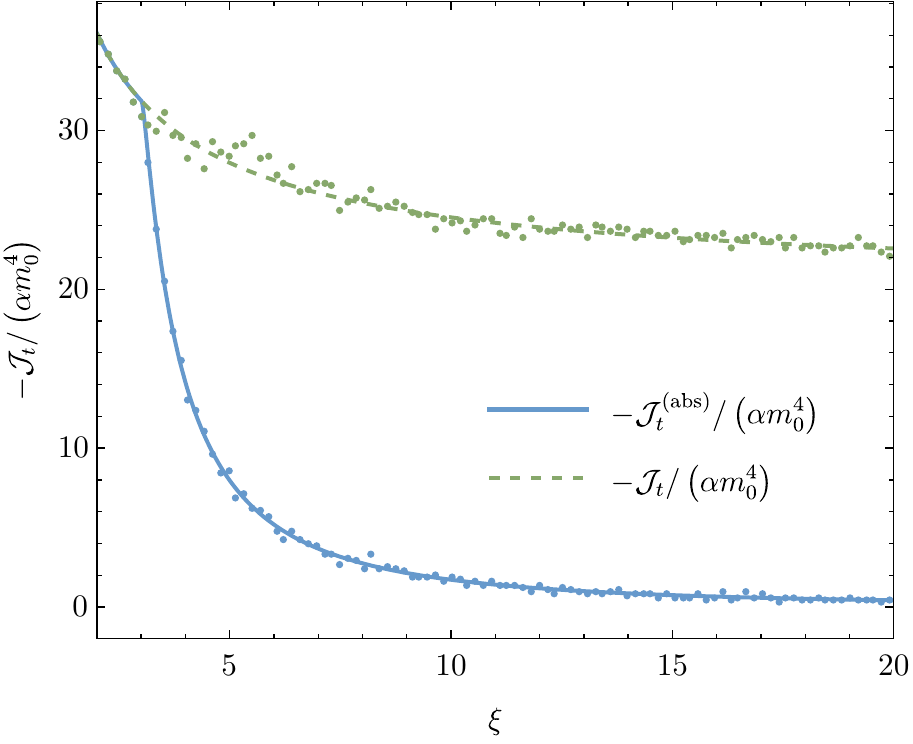}
    \caption{Time components of the particle current density $\mathcal J_t$ for the Maxwell-J\"{u}ttner spherically symmetric model with $\beta = 1$, $\varepsilon_\mathrm{cutoff} = 10$, and $\xi_0 = 20$. Exact solutions (Eqs.\ (\ref{Jt3DexpEqabs}) and (\ref{Jt3DexpEqscat})) are plotted with solid and dashed lines. Blue and green dots show results of a Monte Carlo simulation (Eqs.\ (\ref{Jtexp})). There are $6 \times 10^6$ sets  $(\varepsilon_i, \lambda_i, \varphi_{0,i}, \epsilon_{\lambda,i}, y_i, z_i)$, giving 963 absorbed trajectories and 12804 scattered trajectories in this simulation.}
    \label{fig:my_label9}
\end{figure}

\begin{figure}
    \centering
    \includegraphics[width=\columnwidth]{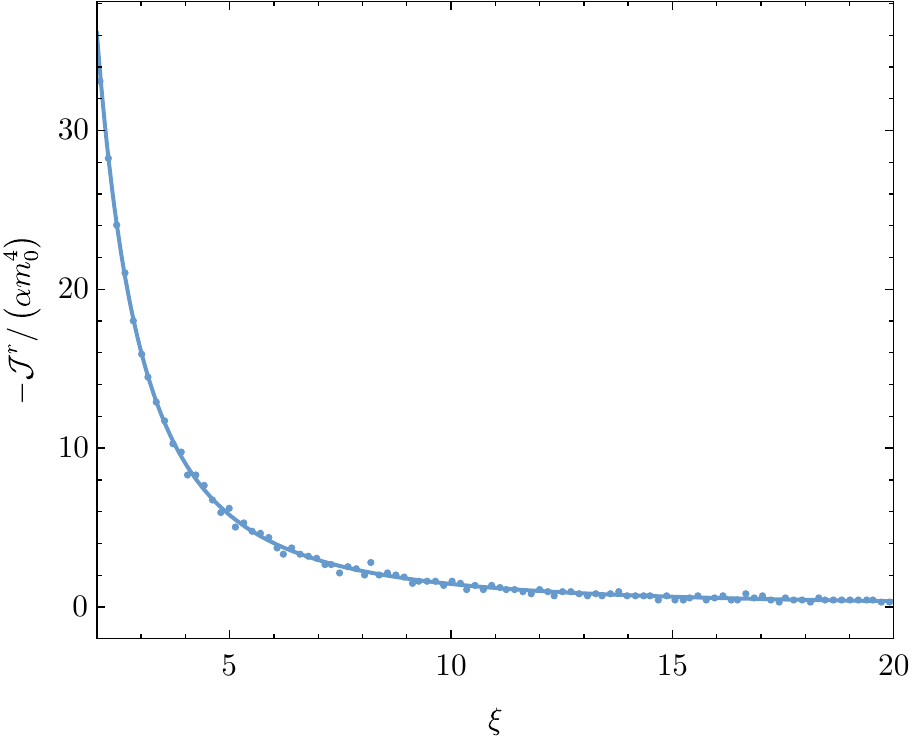}
    \caption{The radial component of the particle current density $\mathcal J^r$ for the Maxwell-J\"{u}ttner spherically symmetric model with $\beta = 1$, $\varepsilon_\mathrm{cutoff} = 10$, and $\xi_0 = 20$. The exact solution (Eq.\ \ref{Jr3DexpEq}) is plotted with the solid line. Blue dots show results of a Monte Carlo simulation (Eqs.\ \ref{Jrexp}). The sample of geodesics orbits is the same as in Fig.\ \ref{fig:my_label9}. There are $6 \times 10^6$ sets  $(\varepsilon_i, \lambda_i, \varphi_{0,i}, \epsilon_{\lambda,i}, y_i, z_i)$, giving 963 absorbed trajectories and 12804 scattered trajectories in this simulation.}
    \label{fig:my_label10}
\end{figure}

\begin{figure}
    \centering
    \includegraphics[width=\columnwidth]{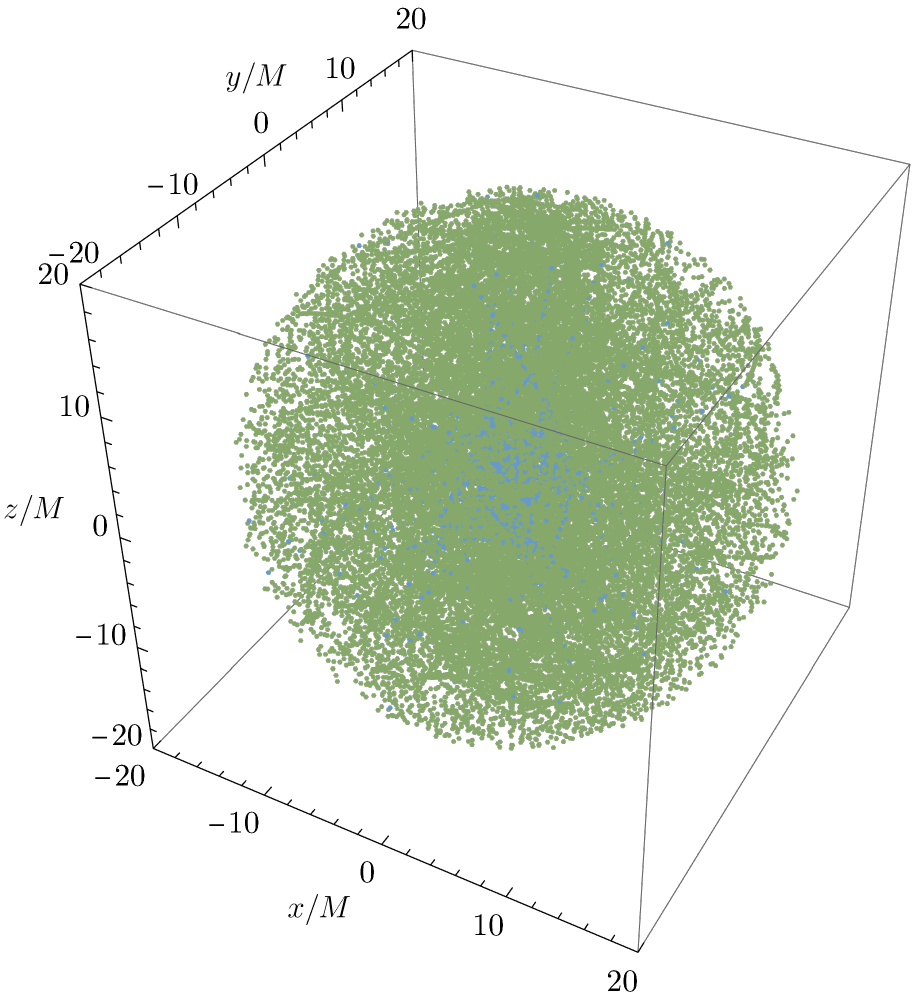}
    \caption{Intersections of trajectories with a grid of meridional half-planes. Blue and green dots correspond to absorbed and scattered orbits, respectively. There are 79 absorbed orbits and 1077 scattered ones.}
    \label{fig:my_label11}
\end{figure}

\begin{figure}
    \centering
    \includegraphics[width=\columnwidth]{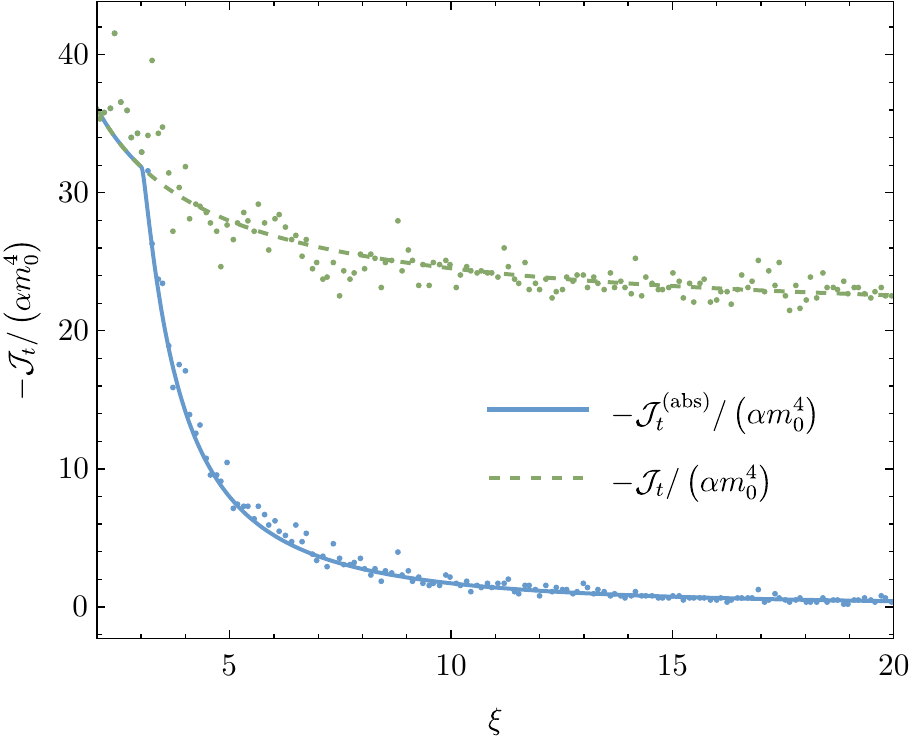}
    \caption{Time components of the particle current density $\mathcal J_t$ for the Maxwell-J\"{u}ttner spherically symmetric model with $\beta = 1$, $\varepsilon_\mathrm{cutoff} = 10$, and $\xi_0 = 20$. Exact solutions (Eqs.\ (\ref{Jt3DexpEqabs}) and (\ref{Jt3DexpEqscat})) are plotted with solid and dashed lines. Blue and green dots show results of a Monte Carlo simulation in which orbital planes of individual trajectories are distributed randomly, and we count intersections of orbits with segments of meridional half-planes of a fixed coordinate system (as shown in Fig.\ \ref{fig:my_label11}). We use von Neumann's rejection method, selecting geodesic parameters and the vector normal to the equatorial plane in a single iteration step. There are $10^7$ iteration steps, giving 828 absorbed and 10979 scattered trajectories.}
    \label{fig:my_label16}
\end{figure}

Similarly to the planar case, one can perform both Monte Carlo integration and the actual Monte Carlo simulation of stationary spherically symmetric accretion flows. Since Monte Carlo integration is essentially straightforward and does not require any new elements, we will omit this part.

We now take
\begin{equation}
S = \{ (r,\theta,\varphi) \colon r_1 \le r \le r_2, \, \theta_1 \le \theta \le \theta_2, \, \varphi = \varphi_0 \}
\end{equation}
and the surface
\begin{eqnarray}
\Sigma & = & \{ (t, r,\theta,\varphi) \colon  t_1 \le t \le t_2, \, r_1 \le r \le r_2, \nonumber \\
&& \theta_1 \le \theta \le \theta_2, \, \varphi = \varphi_0 \}
\end{eqnarray}
and compute the estimators
\begin{equation}
\langle \mathcal J_\mu \rangle = \frac{\int_\Sigma \mathcal J_\mu \eta_\Sigma}{\int_\Sigma \eta_\Sigma}.
\end{equation}
The calculation is analogous to the previous one, performed for the planar case. The expression for the particle current density reads
\begin{equation}
\mathcal J_\mu(x) = \int \sum_{i = 1}^N \delta^{(4)} \left( x^\alpha - x^\alpha_{(i)}  (\tau) \right) \frac{p_\mu^{(i)}(\tau)}{r^2 \sin \theta} d \tau.
\end{equation}
The volume element on $\Sigma$ can be written as $\eta_\Sigma = r dt dr d \theta$. Thus
\begin{equation}
\int_\Sigma \eta_\Sigma = \frac{1}{2}(t_2 - t_1)(r_2 - r_1)(\theta_2 - \theta_1)(r_1 + r_2), \end{equation}
and
\begin{equation}
\int_\Sigma \mathcal J_\mu \eta_\Sigma =  \sum_{i = 1}^{N_\mathrm{int}} \frac{p_\mu^{(i)} r_{(i)} \sin \theta_{(i)}}{|l_z^{(i)}|}.
\end{equation}
The values $l_z$ and $l$ are related: $|l_z| = l \cos \iota$, where $\iota$ denotes the inclination of the orbit with respect to the equatorial plane of the coordinate system. Thus
\begin{equation}
\int_\Sigma \mathcal J_\mu \eta_\Sigma =  \sum_{i = 1}^{N_\mathrm{int}} \frac{p_\mu^{(i)} r_{(i)} \sin \theta_{(i)}}{l_{(i)} \cos \iota_{(i)}}.
\end{equation}
This gives the Monte Carlo estimator of $\mathcal J_\mu$ in the form
\begin{eqnarray}
\langle \mathcal J_\mu \rangle & = & \frac{1}{(t_2 - t_1)(r_2 - r_1)(\theta_2 - \theta_1)} \sum_{i = 1}^{N_\mathrm{int}} \frac{p_\mu^{(i)} \sin \theta_{(i)}}{l_{(i)} \cos \iota_{(i)}} \nonumber \\
& = & \frac{1}{M^2 m (t_2 - t_1)(\xi_2 - \xi_1)(\theta_2 - \theta_1)} \nonumber \\
&& \times \sum_{i = 1}^{N_\mathrm{int}} \frac{p_\mu^{(i)} \sin \theta_{(i)}}{\lambda_{(i)} \cos \iota_{(i)}}
\label{sphsymgeneral}
\end{eqnarray}
where we have assumed that $r_2 - r_1 \ll 1$, and, consequently, $\frac{1}{2}(r_1 + r_2) \approx r_{(i)}$.

The easiest way of simulating spherically symmetric flows in the Schwarzschild background uses the fact that each trajectory is in fact a planar one. Thus, one can still select parameters of trajectories in a single plane. In this case, one can ignore the factors depending on $\theta$ and the inclination angle, and write
\begin{equation}
\langle \mathcal J_\mu \rangle \propto \frac{1}{(t_2 - t_1)(\xi_2 - \xi_1)} \sum_{i = 1}^{N_\mathrm{int}} \frac{p_\mu^{(i)}}{\lambda_{(i)}}.
\end{equation}

The key difference with respect to the planar case is related to the selection of geodesics. While in the planar case we select geodesics from the distribution $\propto d \lambda$, the three-dimensional case requires a distribution $\propto \lambda d \lambda$. To see this, one can once again invoke to a classic discussion of a uniform random distribution of straight lines in the three-dimensional flat space given by Kendall and Moran in \cite{kendall}. They parametrize straight lines in $\mathbb R^3$ by
\begin{subequations}
\label{kmline}
\begin{eqnarray}
    x & = & az + p, \\
    y & = & bz + q,
\end{eqnarray}
\end{subequations}
where $(x,y,z)$ denote Cartesian coordinates. The appropriate probability measure, invariant with respect to Euclidean rotations and translations, is given by
\begin{equation}
(1 + a^2 + b^2)^{-2} da db dp dq
\end{equation}
(cf. Eq.\ (3.50), p.\ 74 in \cite{kendall}). The plane perpendicular to the line given by Eqs.\ (\ref{kmline}) and passing through the origin of the coordinate system is given by
\begin{equation}
\label{kmplane}
ax + by + z = 0.
\end{equation}
The unit vector normal to this plane has the coordinates
\begin{equation}
    (n_x,n_y,n_z) = \frac{(a,b,1)}{\sqrt{1 + a^2 + b^2}}.
\end{equation}
Let us fix the parameters $a$ and $b$. The intersection of the line (\ref{kmline}) with the plane (\ref{kmplane}) is specified by the parameters $p$ and $q$. The area element at the plane (\ref{kmplane}) defined by such intersection points and obtained by varying the parameters $p$ and $q$ is given by
\begin{equation}
    dS = (1 + a^2 + b^2)^{-\frac{1}{2}} dp dq
\end{equation}
(note an error in \cite{kendall} in this formula). Let $d \Omega$ denote the solid angle element around the normal vector $n^i$, obtained by varying $a$ and $b$. Kendall and Moran show that
\begin{equation}
    (1 + a^2 + b^2)^{-2} da db dp dq = dS d\Omega.
\end{equation}

Let $\delta$ be a distance from the center of the coordinate system to the intersection point. It is given by
\begin{equation}
    \delta = \sqrt{\frac{(1 + b^2)p^2 - 2 a b p q + (1 + a^2) q^2}{1 + a^2 + b^2}}.
\end{equation}
Fix the parameters $a$ and $b$. Let $\phi$ denote an angle between the intersection of the plane (\ref{kmplane}) and the plane $z = 0$ and the line joining the origin of the coordinate system and the intersection point. We have
\begin{equation}
\cos \phi = \frac{aq - bp}{\delta \sqrt{a^2 + b^2}}.
\end{equation}
One can show that
\begin{equation}
\frac{\partial (\delta,\phi)}{\partial(p,q)} = \pm \frac{1}{\delta \sqrt{1 + a^2 + b^2}}.
\end{equation}
Thus,
\begin{equation}
dS = \delta d\delta d \phi,
\end{equation}
as expected ($(\delta, \phi)$ are standard polar coordinates in the plane (\ref{kmplane})). Again, the total angular momentum $\lambda$ associated with a given line is directly proportional to its distance $\delta$ from the center of the coordinate system. This justifies our claim that
\begin{equation}
    (1 + a^2 + b^2)^{-2} da db dp dq \propto \lambda d \lambda.
\end{equation}

Assuming the notation from the previous section, we write Monte Carlo estimators for the monoenergetic model in the following form:
\begin{subequations}
\label{Jtmono}
\begin{eqnarray}
\langle \mathcal J_t^\mathrm{(abs)} \rangle & = &- \frac{2 \pi^2 \alpha m_0^4 \lambda_c(\varepsilon_0)^2}{N_\mathrm{abs} (\xi_2 - \xi_1)} \sum_{i \in I_\mathrm{abs}(\xi_1,\xi_2)} \frac{\varepsilon_0}{\lambda_{(i)}}, \\
\langle \mathcal J_t^\mathrm{(scat)} \rangle & = & - \frac{2 \pi^2 \alpha m_0^4 [ \lambda_\mathrm{max}(\varepsilon_0, \xi_0)^2 -\lambda_c(\varepsilon_0)^2]}{N_\mathrm{scat} (\xi_2 - \xi_1)}  \nonumber \\
&& \times \sum_{i \in I_\mathrm{scat}(\xi_1,\xi_2)} \frac{\varepsilon_0}{\lambda_{(i)}}.
\end{eqnarray}
\end{subequations}
The estimator for $\mathcal J^r$ can be computed as
\begin{equation}
\label{Jrmono}
\langle \mathcal J^r \rangle = - \frac{2 \pi^2 \alpha m_0^4 \lambda_c(\varepsilon_0)^2}{N_\mathrm{abs} (\xi_2 - \xi_1)} \sum_{i \in I_\mathrm{abs}(\xi_1,\xi_2)} \frac{\sqrt{\varepsilon_0^2 - U_{\lambda_{(i)}}\left(\xi_{(i)}\right)}}{\lambda_{(i)}}.
\end{equation}
Here, as in the planar case, we have adapted our normalization to analytic formulas (\ref{Jt3DmonoEq}) and (\ref{Jr3DmonoEq}). A comparison of these estimators with the exact solutions is given in Fig.\ \ref{fig:my_label7} and \ref{fig:my_label8}.

In a similar fashion one can compute Monte Carlo estimators of the particle current density in the case with the Maxwell-J\"{u}ttner asymptotic distribution. As in the planar case, parameters of the trajectories can be selected from the Maxwell-J\"{u}ttner distribution, using von Neumann's rejection method. Once again, we repeat the procedure described in Sec.\ \ref{sec:montecarlointegration}. The only difference is that this time both angular momenta $\lambda$ and the energies $\varepsilon$ are selected from non-uniform distributions. This requires introducing two separate auxiliary parameters ($y_i$ and $z_i$, say). Monte Carlo estimators of $\mathcal J_t^\mathrm{(abs)}$ and $\mathcal J_t^\mathrm{(scat)}$ can be computed as
\begin{subequations}
\label{Jtexp}
\begin{eqnarray}
\langle \mathcal J_t^\mathrm{(abs)} \rangle & = &- \frac{4 \pi^2 \alpha m_0^4 \hat V_\mathrm{abs}}{N_\mathrm{abs} (\xi_2 - \xi_1)}  \sum_{i \in I_\mathrm{abs}(\xi_1,\xi_2)} \frac{\varepsilon_{(i)}}{\lambda_{(i)}}, \\
\langle \mathcal J_t^\mathrm{(scat)} \rangle & = & - \frac{4 \pi^2 \alpha m_0^4 \hat V_\mathrm{scat}}{N_\mathrm{scat} (\xi_2 - \xi_1)} \sum_{i \in I_\mathrm{scat}(\xi_1,\xi_2)} \frac{\varepsilon_{(i)}}{\lambda_{(i)}}.
\end{eqnarray}
\end{subequations}
The estimator for $\mathcal J^r$ reads
\begin{equation}
\label{Jrexp}
\langle \mathcal J^r \rangle = - \frac{4 \pi^2 \alpha m_0^4 \hat V_\mathrm{abs}}{N_\mathrm{abs} (\xi_2 - \xi_1)} \sum_{i \in I_\mathrm{abs}(\xi_1,\xi_2)} \frac{\sqrt{\varepsilon_{(i)}^2 - U_{\lambda_{(i)}}\left(\xi_{(i)}\right)}}{\lambda_{(i)}}.
\end{equation}
Here
\begin{subequations}
\label{volumes2}
\begin{eqnarray}
\hat V_\mathrm{abs} & = &  \frac{1}{2} \int_1^{\varepsilon_\mathrm{cutoff}} \exp(- \beta \varepsilon) \lambda_c(\varepsilon)^2 d \varepsilon, \\
\hat V_\mathrm{scat} & = & \frac{1}{2} \int_1^{\varepsilon_\mathrm{cutoff}} \exp(- \beta \varepsilon) \nonumber \\
&& \times [\lambda_\mathrm{max}(\varepsilon,\xi_0)^2 - \lambda_c(\varepsilon)^2] d \varepsilon.
\end{eqnarray}
\end{subequations}
A comparison of Monte Carlo estimators defined in this way and exact solutions is given in Figs.\ \ref{fig:my_label9} and \ref{fig:my_label10}.

Of course, one can also adhere to the original prescription given by Eq.\ (\ref{sphsymgeneral}) and distribute selected trajectories among randomly oriented orbital planes. This requires some technical elements in the calculation, but does not change overall results. The idea is to count (with appropriate weights) intersections of trajectories with meridian half-planes of a fixed spherical coordinate system. As before, we count separately intersections falling in the radial regions $\xi_j < \xi < \xi_{j+1}$, $j = 1, \dots, N_\xi - 1$, with $\xi_1 = 2$ and $\xi_{N_\xi} = \xi_0$, but we also discretize with respect to $\theta$, and count separately intersections occurring in different regions $\theta_k < \theta < \theta_{k+1}$, $k = 1, \dots, N_\theta - 1$, $\theta_1 = 0$ and $\theta_{N_\theta} = \pi$. As before, for clarity, we keep the notation with $\xi_1$, $\xi_2$, $\theta_1$, and $\theta_2$ referring to cell boundaries.

The Monte Carlo estimators are then computed as
\begin{widetext}
\begin{subequations}
\label{Jtexp3Dtrue}
\begin{eqnarray}
\langle \mathcal J_t^\mathrm{(abs)} \rangle & = &- \frac{8 \pi \alpha m_0^4 \hat V_\mathrm{abs}}{N_\mathrm{abs} (\xi_2 - \xi_1)(\theta_2 - \theta_1)} \sum_{i \in I_\mathrm{abs}(\xi_1, \xi_2;\theta_1, \theta_2)} \frac{\varepsilon_{(i)} \sin \theta_{(i)}}{\lambda_{(i)} \cos \iota_{(i)}}, \\
\langle \mathcal J_t^\mathrm{(scat)} \rangle & = & - \frac{8 \pi \alpha m_0^4 \hat V_\mathrm{scat}}{N_\mathrm{scat} (\xi_2 - \xi_1)(\theta_2 - \theta_1)} \sum_{i \in I_\mathrm{scat}(\xi_1, \xi_2; \theta_1, \theta_2)} \frac{\varepsilon_{(i)} \sin \theta_{(i)}}{\lambda_{(i)} \cos \iota_{(i)}}, \\
\langle \mathcal J^r \rangle & = &- \frac{8 \pi \alpha m_0^4 \hat V_\mathrm{abs}}{N_\mathrm{abs} (\xi_2 - \xi_1)(\theta_2 - \theta_1)} \sum_{i \in I_\mathrm{abs}(\xi_1, \xi_2;\theta_1, \theta_2)} \frac{\sqrt{\varepsilon_{(i)}^2 - U_{\lambda_{(i)}}\left(\xi_{(i)}\right)} \sin \theta_{(i)}}{\lambda_{(i)} \cos \iota_{(i)}},
\end{eqnarray}
\end{subequations}
\end{widetext}
where $I_\mathrm{abs}(\xi_1, \xi_2;\theta_1, \theta_2)$ and $I_\mathrm{scat}(\xi_1, \xi_2; \theta_1, \theta_2)$ collect indices corresponding to \textit{intersections} of absorbed and scattered trajectories falling in the regions $\xi_1 \le \xi_{(i)} \le \xi_2$ and $\theta_1 \le \theta_{(i)} \le \theta_2$.

The orientation of the orbital plane can be controlled by specifying coordinates of the normal vector. Choosing to work with Cartesian coordinates facilitates the selection procedure---again von Neumann's rejection method can be used to select normal vectors distributed uniformly in a unit sphere. Taking into account different possible orientations of the orbits would be important for non-spherically symmetric solutions. Examples of such solutions, representing models of a Schwarzschild black hole moving through the medium, are provided in \cite{pmao1, pmao2}. For spherically symmetric solutions, one can average the results not only over different values of $\varphi_0$, but also over cells $[\theta_1, \theta_2]$.

For completeness, Fig.\ \ref{fig:my_label11} shows an example of intersections of orbits selected according to the Maxwell-J\"{u}ttner distribution with a grid of meridional half-planes. Sample results of a Monte Carlo simulation with a fixed coordinate system and randomly distributed orbital planes of individual trajectories are shown in Fig.\ \ref{fig:my_label16}.

\section{Summary}
\label{sec:summary}

We have developed Monte Carlo techniques of computing stationary solutions of the general-relativistic Vlasov equation. For simplicity, we worked with stationary accretion-type solutions on the Schwarzschild background, but with appropriate adjustments, our methods should also work in other cases, including stationary configurations of the collisionless gas in the Kerr spacetime. (A possible future implementation valid for the Kerr spacetime should, in particular, recover results of the planar model \cite{cieslik_mach_odrzywolek_2022} as a test.)

Our discussion emphasised the difference between planar models (in which the motion of particles is confined to a common plane) and non-planar ones. It is especially tricky (or subtle) for spherically symmetric spacetimes (like the Schwarzschild spacetime), in which each of the geodesics belongs to a single plane.

In our examples, we have concentrated on computing the particle current density, but other observable quantities can be obtained in a similar fashion. A simple Monte Carlo simulation yielding the rest-mass accretion rate in the low-temperature limit of the Bondi-Hoyle-Lyttleton type models described in \cite{pmao1, pmao2} was reported in \cite{pmao3}.

Aside from a purely technical aspect of the proposed Monte Carlo method, we hope that the cases investigated in this paper offer an insight in the geometric structure of the general-relativistic kinetic theory and its statistical interpretation.

\begin{acknowledgments}
We are grateful to Olivier Sarbach for discussions. This work was partially supported by the Polish National Science Centre Grant No.\ 2017/26/A/ST2/00530.
\end{acknowledgments}

\end{document}